\documentclass[twocolumn,preprintnumbers,amsmath,amssymb,superscriptaddress,nofootinbib,longbibliography, prb]{revtex4-2}

\usepackage{graphicx}
\usepackage{bm}
\usepackage{dsfont}
\usepackage[usenames,dvipsnames]{xcolor}
\usepackage{pstricks}
\usepackage[tight]{subfigure}
\usepackage{verbatim}
\usepackage{units}
\usepackage{multirow}
\usepackage{mathrsfs}
\usepackage{leftidx}
\usepackage{xspace}
\usepackage{diffcoeff}  
\usepackage{bbm}
\usepackage{amsfonts,amssymb,amsmath}
\usepackage{mathtools}

\usepackage{array} 
\usepackage{tabularx}
\usepackage[normalem]{ulem}
\usepackage{bbold}
\usepackage{pifont}
\usepackage{hyperref}
\hypersetup{colorlinks=true,linktoc=all,linkcolor=blue,breaklinks=true,citecolor=blue,urlcolor=blue}

\usepackage[capitalise]{cleveref}

\usepackage[utf8]{inputenc}
\usepackage[english]{babel}

\definecolor{darkgreen}{RGB}{6, 153, 38}

\addto\captionsenglish{\renewcommand{\figurename}{FIG.}}

\AtBeginDocument{%
    \newwrite\bibnotes
    \def\bibnotesext{Notes.bib}
    \immediate\openout\bibnotes=\jobname\bibnotesext
    \immediate\write\bibnotes{@CONTROL{REVTEX42Control}}
    \immediate\write\bibnotes{@CONTROL{%
    apsrev42Control,author="08",editor="1",pages="1",title="0",year="1"}}
     \if@filesw
     \immediate\write\@auxout{\string\citation{apsrev42Control}}%
    \fi
}%

\newcommand*{\Tr}{\text{Tr}}

\newcommand{\beq}{\begin{equation}}
\newcommand{\eeq}{\end{equation}}

\DeclarePairedDelimiterX\braket[2]{\langle}{\rangle}{#1\,\delimsize\vert\,\mathopen{}#2}
\DeclarePairedDelimiterX\ketbra[2]{\lvert}{\rvert}{#1\,\delimsize\rangle\mathopen{}\delimsize\langle\,\mathopen{}#2}

\DeclarePairedDelimiterX\Braket[2]{(}{)}{#1\,\delimsize\vert\,\mathopen{}#2}
\DeclarePairedDelimiterX\Ketbra[2]{\lvert}{\rvert}{#1\,\delimsize)\mathopen{}\delimsize(\,\mathopen{}#2}

\newcommand{\kB}{k_\text{B}}
\newcommand{\kBT}{k_\text{B}T}

\newcommand{\md}{\mathrm{d}}
\newcommand{\mi}{\mathrm{i}}

\usepackage{sidecap,tikz}
\definecolor{lime}{HTML}{A6CE39}
\DeclareRobustCommand{\orcidicon}{\hspace{-1mm}
	\begin{tikzpicture}
		\draw[lime, fill=lime] (0,0) 
		circle [radius=0.16] 
		node[white] {{\fontfamily{qag}\selectfont \tiny \,ID}};
		\draw[white, fill=white] (-0.0525,0.095) 
		circle [radius=0.007];
	\end{tikzpicture}
	\hspace{-3mm}
}
\foreach \x in {A, ..., Z}{\expandafter\xdef\csname orcid\x\endcsname{\noexpand\href{https://orcid.org/\csname orcidauthor\x\endcsname}
		{\noexpand\orcidicon}}
}

\begin{document}

\title{
Thermoelectric processes of quantum normal-superconductor interfaces
}
\author{Liliana Arrachea\orcidA{}}
\affiliation{Centro At\'omico Bariloche, Instituto de Nanociencia y Nanotecnolog\'ia CONICET-CNEA and Instituto Balseiro (8400), San Carlos de Bariloche, Argentina\looseness=-1}
\author{Alessandro Braggio\orcidB{}}
\affiliation{NEST, Istituto Nanoscienze-CNR and Scuola Normale Superiore, Piazza San Silvestro 12, I-56127 Pisa, Italy\looseness=-1}
\affiliation{Institute for Quantum Studies, Chapman University, Orange, CA 92866, USA
}
\author{Pablo Burset\orcidC{}}
\affiliation{Departamento de F\'isica Te\'orica de la Materia Condensada, Universidad Aut\'onoma de Madrid, 28049 Madrid, Spain\looseness=-1}
\affiliation{Condensed Matter Physics Center (IFIMAC), Universidad Aut\'onoma de Madrid, 28049 Madrid, Spain\looseness=-1}
\affiliation{Instituto Nicol\'as Cabrera (INC), Universidad Aut\'onoma de Madrid, 28049 Madrid, Spain\looseness=-1}
\author{Eduardo J. H. Lee\orcidD{}}
\affiliation{Departamento de F\'isica de la Materia Condensada, Universidad Aut\'onoma de Madrid, 28049 Madrid, Spain\looseness=-1}
\affiliation{Condensed Matter Physics Center (IFIMAC), Universidad Aut\'onoma de Madrid, 28049 Madrid, Spain\looseness=-1}
\affiliation{Instituto Nicol\'as Cabrera (INC), Universidad Aut\'onoma de Madrid, 28049 Madrid, Spain\looseness=-1}
\author{Alfredo Levy Yeyati\orcidE{}}
\affiliation{Departamento de F\'isica Te\'orica de la Materia Condensada, Universidad Aut\'onoma de Madrid, 28049 Madrid, Spain\looseness=-1}
\affiliation{Condensed Matter Physics Center (IFIMAC), Universidad Aut\'onoma de Madrid, 28049 Madrid, Spain\looseness=-1}
\affiliation{Instituto Nicol\'as Cabrera (INC), Universidad Aut\'onoma de Madrid, 28049 Madrid, Spain\looseness=-1}
\author{Rafael S\'anchez\orcidF{}}
\affiliation{Departamento de F\'isica Te\'orica de la Materia Condensada, Universidad Aut\'onoma de Madrid, 28049 Madrid, Spain\looseness=-1}
\affiliation{Condensed Matter Physics Center (IFIMAC), Universidad Aut\'onoma de Madrid, 28049 Madrid, Spain\looseness=-1}
\affiliation{Instituto Nicol\'as Cabrera (INC), Universidad Aut\'onoma de Madrid, 28049 Madrid, Spain\looseness=-1}
\date{\today}

\begin{abstract}
Superconducting interfaces have recently been demonstrated to contain a rich variety of effects that give rise to sizable thermoelectric responses and unexpected thermal properties, despite traditionally being considered poor thermoelectrics due to their intrinsic electron-hole symmetry. 
We review different mechanisms driving this response in hybrid normal-superconducting junctions, depending on the dimensionality of the mesoscopic interface. In addition to discussing heat to power conversion, cooling and heat transport, special emphasis is put on physical properties of hybrid devices that can be revealed by the thermoelectric effect.
\end{abstract}

\maketitle

\section{Introduction}
The research on thermoelectricity has been mainly developed from the perspective of heat harvesting and energy applications~\cite{Bell2008Sep,Snyder2008Feb,Heremans2013Jul,Zhang2015Jun}. However, with recent developments in mesoscopic physics~\cite{Imry2001Dec,Altshuler1991,Datta1995Sep}, these peculiar transport characteristics have sparked new interest~\cite{Sivan1986Jan,Butcher1990Jun,Amman1991Oct,Guttman1995Jun}. The main reason is not simply due to power generation and improving 
efficiency~\cite{Hicks1993May,Hicks1993Jun,Mahan1996Jul,Whitney2014Apr} but more intriguingly as a method to potentially detect novel physical properties~\cite{Kuzanyan2012Mar,Nam2013May,Lundeberg2017Feb,Heikkila2018Sep,Kleeorin2019Dec,Guarcello2023Sep}. In recent years, the development of the field of quantum thermodynamics  
and the advent of superconducting-based quantum technologies~\cite{Aguado2020Dec} have posed an interest in exploring those properties in hybrid superconducting mesoscopic devices. 
The possibility to develop novel experimental platforms where thermal imbalances are maintained in mesoscopic regimes~\cite{Giazotto2006Mar,Pekola2021Oct} or are generated by the transport further motivates the investigation of thermoelectrical effects in such systems~\cite{Benenti2017Jun,Balduque2025Apr}. 
Charge and heat transport~\cite{Oettinger2014Oct} and thermoelectric properties in 
mesoscopic conductors~\cite{Heikkila2013Jan} with superconductors have been previously discussed in many papers and reviews  \cite{Claughton1996Mar,Virtanen2007Nov,DeFranceschi2010Oct,Lemziakov2024Oct,Dutta2025Mar} and we refer the interested reader to them. 

\begin{figure}[b]
  \includegraphics[width=\linewidth]{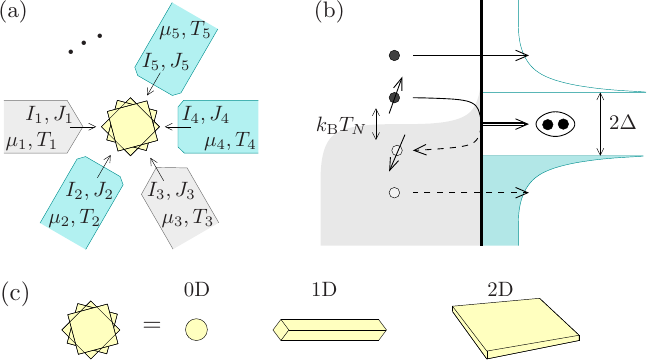}
  \caption{Scope of the review. (a) Scheme of a hybrid configuration composed of normal (gray) and superconducting (cyan) terminals separated by a mesoscopic region (yellow). Each terminal $l$ is in a local equilibrium situation defined by its chemical potential, $\mu_l$, and temperature, $T_l$, resulting in particle, $I_l$, and heat, $J_l$, currents. (b) Transport occurs due to quasiparticle (electron or hole-like) tunneling or to Andreev reflection, here represented by the creation of a Cooper pair in the superconductor. (c) The mesoscopic region can be of different dimensionalities: 0D (e.g., quantum dots), 1D (e.g., quantum wires, topological edge states), 2D (e.g., graphene, semiconductor 2DEGs, surfaces of 3D topological insulators), depending on the properties of the propagating particles.}
  \label{fig:processes}
\end{figure}

In this review, we will concentrate on recent examples where the thermoelectric response crucially characterizes the transport in hybrid systems composed of normal metal (N) and superconducting (S) terminals connected by a mesoscopic (low-dimensional) structure, as sketched in Fig.~\ref{fig:processes}. 
We characterize different platforms spanning from quantum dots \cite{Sothmann2014Dec,Sanchez2018Dec,Hussein2019Feb}, nanowires~\cite{Lee2014Jan,Prada2020Oct}, helical modes~\cite{Mateos2024Aug}, topological materials, both 2D and 3D, which are proximized or coupled to superconducting materials.
The thermoelectric effect in hybrid superconducting systems has salient properties in these geometries, which have been discussed for different reasons, from the detection of helical states~\cite{Blasi2020Jun,Blasi2021Jun} and Majorana bound states~\cite{Leijnse2014Jan,Lopez2014May,Benjamin2024Apr,Klees2024Aug}, to odd-frequency superconductivity~\cite{Hwang2018Oct} or order parameter symmetries~\cite{Savander2020Dec,Guarcello2023Sep}.

It is instructive to recall that the existence of thermoelectricity in bulk superconductors was firstly predicted in a series of seminal works by Ginzburg~\cite{Ginzburg1944,Ginzburg1978May,Ginzburg1991Jan}, which have motivated experimental research~\cite{VanHarlingen1982Jul,Shelly2016Feb}.
Thermoelectric~\cite{Guttman1997May} and thermal~\cite{Maki1965Dec,Guttman1997Feb,Guttman1998Feb} transport in Josephson junctions induced by quasiparticle-Cooper pair coupling have been measured~\cite{Giazotto2012Dec} and exploited in a number of {\it coherent caloritronic} devices~\cite{Martinez-Perez2014Jun,Hwang2020Feb}. Recently, strong thermoelectrical effects have been reported in superconducting tunnel junctions done with ferromagnetic superconductors~\cite{Heikkila2013Jan,Machon2013Jan,Ozaeta2014Feb,Kolenda2016Mar}, in SIS (I being an insulating barrier) junctions with asymmetric gap configurations~\cite{Marchegiani2020Mar,Germanese2022Oct}, magnetic molecules~\cite{Volosheniuk2025Apr}, or even Abrikosov vortices~\cite{Singh2024Dec}. Intriguingly, phase-dependent thermoelectric effects have also been discussed in Andreev interferometers, where good agreement between semiclassical theory in diffusive regimes ~\cite{Seviour2000Sep,Virtanen2004Apr,Virtanen2007Mar,Titov2008Dec} or 
scattering theory for mesoscopic diffusive~\cite{Claughton1996Mar} or chaotic cavities~\cite{Jacquod2010Oct,Engl2011May} have been contrasted with experiments~\cite{Eom1998Jul,Parsons2003Apr,Jiang2005Jul}.
In this review, we mainly concentrate on hybrid configurations between normal, topological materials and superconductors, and hence we will not discuss all-superconductor or superconductor-ferromagnetic interfaces, referring the reader to specific reviews~\cite{Bergeret2018Oct,Cangemi2024Oct, Dutta2025Mar}.

\section{Physical processes}
\label{sec:physics}
Thermoelectricity in mesoscopic systems has been reviewed in detail in Ref.~\onlinecite{Benenti2017Jun} and the expected consequences of the different symmetry classes on the linear thermoelectric coefficients have been discussed in Ref.~\onlinecite{Jacquod2012Oct}. In the presence of one or more superconducting terminals, we refer the reader to the well-known literature~\cite{Claughton1996Mar,Lambert1993Jun,Lambert1998Feb}. 
We now briefly review the basic results. 

The Bogoliubov de Gennes (BdG) formalism~\cite{DeGennes2018Mar} in combination with a scattering picture of coherent electronic transport provides a description of the most basic phenomena. A typical model Hamiltonian for a hybrid nanostructure is
\begin{equation}
H = \frac{1}{2} \int \md r \hat{\psi}^{\dagger}(r) {\cal H}_{\rm BdG}(r) \hat{\psi}(r) ,
\end{equation}
where $\hat{\psi}^T(r) = \left( \psi_{\uparrow}(r), \; \psi_{\downarrow}(r), \; \psi^{\dagger}_{\downarrow}(r), \; -\psi^{\dagger}_{\uparrow}(r) \right)$ is a field Nambu spinor acting on spin and particle-hole space, and
\begin{equation}\label{eq:BdG}
    {\cal H}_{\rm BdG} = \left( \begin{array}{cc} \hat{h}_e(r) - \mu & \hat{\Delta}(r) \\
    \hat{\Delta}^{\dagger}(r) & - \mathbf{\hat{T}}\left[\hat{h}_{e}(r) - \mu\right]\mathbf{\hat{T}}^{-1} \end{array}\right) ,
\end{equation}
with $\hat{h}_e(r)$ and $\mu$ denoting the electronic Hamiltonian and the chemical potential, respectively, while $\mathbf{\hat{T}}$ is the time-reversal operator. The pairing amplitude $\hat{\Delta}(r)$ couples electrons and holes in the superconductors and vanishes outside them. In semiconductor-superconductor junctions, in the low-energy limit, $\hat{\Delta}(r)$ is usually modeled as a step-function~\cite{Beenakker1997Jul}. 

Superconducting electron-hole correlations generated at the interface due to Andreev reflections~\cite{Andreev1964May} at subgap energies determine the superconducting proximity effect, which induces a finite pairing amplitude penetrating into the normal regions. Most interestingly, the Andreev processes entail finite scattering amplitudes for incident electrons to be reflected as holes into the same lead or transmitted into another normal terminal, ${\cal S}^{he}$, or vice versa, ${\cal S}^{eh}$, see later. 
These processes are correspondingly accompanied by a Cooper pair being either created or annihilated in the superconducting condensate, as sketched in Fig.~\ref{fig:processes}(b). It is worth noting that the Andreev reflection mechanism was originally proposed to explain thermal transport in N-S junctions~\cite{Andreev1964May}, with the superconductor acting as a mirror for heat in this regime.

In general, for subgap energies $|E|<\Delta$, there are no propagating states in the bulk of a superconductor and Andreev reflection is local (LAR) on a simple NS interface. 
However, for superconductors of size comparable to the coherence length $\xi$, those states still exist in the form of evanescent waves which could mediate superconducting cross-correlations at finite distances inside the superconductors. This results in crossed (nonlocal) Andreev reflections (CAR). 
The interplay between LAR, CAR and standard scattering properties frequently characterizes hybrid mesoscopic superconducting nanodevices, sometimes determining unexpected thermo-electrical phenomena, as we will see. 

Using scattering theory~\cite{Anantram1996Jun,Blanter2000Sep,Heikkila2013Jan}, transport is then understood in terms of the amplitudes for electrons ($\alpha=e$, $s_e=+1$) or holes ($\alpha=h$, $s_h=-1$) incident with energy $s_\alpha E$ from channel $m$ of one normal terminal, $l$, to be scattered into channel $m'$ of another terminal, $l'$, ${\cal S}^{\alpha'\alpha}_{m'l',ml}(E)$~\cite{Beenakker1992Nov,Anantram1996Jun,Beenakker1997Jul}. 
Recall that the pairing amplitude in the superconducting regions make ${\cal S}^{he}_{m'l',ml^{}}$ and ${\cal S}^{eh}_{m'l',ml^{}}$ finite, resulting in LAR, when $l'=l^{}$, or CAR~\cite{Byers1995Jan,Anantram1996Jun,Deutscher2000Jan}, when $l'\neq l^{}$. In the subgap regime, ${\cal S}^{\alpha\alpha}_{m'l',ml^{}}$ potentially also includes contributions due to the elastic cotunneling (EC) of a particle via the superconductor~\cite{Falci2001Apr}. The nonlocal amplitudes with $l'\neq l^{}$ between the two normal leads (coupled via the superconductor), as discussed above, are usually limited to terminals separated by a distance shorter than $\xi$. 

Increasing the energies (voltages or temperatures), other normal (non-Andreev) processes emerge which do not involve an exchange of Cooper pairs with the condensate. 
An extreme case is when the NS interface is very opaque (tunneling limit) so that superconducting correlations between the two terminals can be ignored. 
As a result, the Andreev scattering amplitudes with $\alpha\neq\alpha'$ become negligible. Scattering between all terminals (including normal and superconducting) is then described only in terms of normal (electron-to-electron or hole-to-hole)  amplitudes ${\cal S}_{m'l',ml}^{\alpha\alpha}$. 
In the following, we will discuss how to connect the scattering matrix to the heat and charge transport in multiterminal superconducting heterostructures.

\subsection{Charge and heat transport}
A general description of transport explicitly respecting the first two laws of thermodynamics~\cite{Benenti2017Jun} is given in terms of scattering theory~\cite{Lambert1993Jun,Anantram1996Jun}. 
The full scattering matrix allows us to calculate transport properties. Specifically, it introduces the scattering probabilities 
\begin{equation}
{\cal T}_{l'l}^{\alpha\beta}(E)=\sum_{mm'}|{\cal S}_{m'l',ml}^{\alpha\beta}(E)|^2 ,
\end{equation}
where the sum is expressed over all $m$($m'$)-modes coming from the the $l$ ($l'$) terminal. 
For simplicity, we limit our discussion to the current in the normal leads, which is given by the sum of the contributions of all the propagating states in the terminal.\footnote{In superconducting terminals, the charge current is the sum of the current from the propagating states plus the dissipationless supercurrent from the condensate.}
In this case, the current carried by $\alpha$-type quasiparticles entering the scatterer from terminal $l$ at a given energy $E$ is 
\begin{equation}
\label{eq:currdensE}
{\cal I}_{l}^\alpha(E)= \frac{1}{2h}\sum_{l',\alpha'}\left[{\cal N}_{l}^\alpha\delta_{l^{}l'}\delta_{\alpha\alpha'}-{\cal T}_{l^{}l'}^{\alpha\alpha'}(E)\right]f_{l'}^{\alpha'}(E),
\end{equation}
where ${\cal N}_{l}^\alpha$ being the number of channels in lead $l$, with the index $l'$ running over all terminals and the sum over $\alpha'$ including both electron and hole states.~\footnote{Note that the sum of $l'$ is done on all the terminals which have asymptotically propagating modes at energy $E$ in the terminal. This means that superconducting terminals need to be properly included in the sum only for $|E|\geq\Delta$.}
The Fermi distributions are $f_l^\alpha(E)=\{1+\exp[(E-s_\alpha\mu_l)/\kB T_l]\}^{-1}$, where $\mu_l$ and $T_l$ are, respectively, the chemical potential and the temperature of terminal $l$. In this simplified treatment, we assume that all the superconducting terminals are kept at the same chemical potential, which is taken as the reference bias, i.e., $\mu_S=0$.
Note that with this unified notation ${\cal T}_{l'l}^{\alpha\beta}$, with $l=l'$ ($l\neq l'$), represents (crossed) Andreev reflection probabilities when $\alpha\neq\beta$, and normal reflections (transmissions) for $\alpha=\beta$.~\footnote{The factor $1/2$ in the operator of Eq.~\eqref{eq:currdensE} comes from the adopted convention to consider states in the Nambu states which extend in energy from $E\in(-\infty,+\infty)$. This means that all the energy integrals, if not differently indicated, have to be defined over that domain. This representation is fully equivalent to the convention restricting only to states in the positive energy domain, $E>0$. The bilateral convention we adopted for specific computations, especially when small temperature differences are considered, such as for linear thermoelectric properties, is more convenient to identify cancellations between terms.}
Summing over the energies in Eq.~\eqref{eq:currdensE}, one can easily compute the total particle and heat currents~\cite{Benenti2017Jun}
\begin{align}
\label{eq:Il}
I_{l^{}}={}& e\sum_\alpha\int{\md E}\ s_\alpha{\cal I}_{l^{}}^\alpha(E) , 
\\
\label{eq:Jl}
J_{l^{}}={}& \sum_\alpha\int{\md E}\ (E-s_\alpha\mu_{l^{}}){\cal I}_{l^{}}^\alpha(E) ,
\end{align}
where $s_\alpha$ keeps track of the opposite contribution of electrons and holes to the charge current in Eq.~\eqref{eq:Il}, and that they are measured differently with respect to the chemical potential in Eq.~\eqref{eq:Jl}.

The currents can also be calculated using standard many-body techniques such as Schwinger-Keldysh non-equilibrium Green's functions based on microscopic Hamiltonians for the normal and superconducting leads (${\cal H}_l$), the mesoscopic region (${\cal H}_{\rm M}$), and tunnel-contacts between them. 
The precise mapping between the scattering theory quantities and the Green function formalism has been discussed in the literature~\cite{Cuevas1996Sep,Atienza2014,Mateos2024Aug}. 
In particular, the transmission functions can be expressed 
in terms of the mesoscopic region retarded/advanced Green function connected to all the leads, 
\begin{equation}
{\bf G}^{r/a}_{\rm M}(E) = \lim\limits_{\eta\to 0^+}[ (E \pm i \eta)\mathbb{1} - {\cal H}_{\rm M} - \boldsymbol{\Sigma}_\text{leads}]^{-1} , 
\end{equation}
with $\mathbb{1}$ the identity matrix and where $\boldsymbol{\Sigma_\text{leads}}$ is the self-energy taking into account the effect of the leads on the mesoscopic region. 
The self-energy due to normal-state leads is diagonal in Nambu space and we can define the hybridization matrices $\boldsymbol{\Gamma}_l^\alpha(E) = 2\mathrm{Im}\{ \boldsymbol{\Sigma}_l^\alpha(E) \} = 2\pi \boldsymbol{\tau}_{{\rm M},l} \boldsymbol{\rho}^\alpha_l(E) \boldsymbol{\tau}_{{\rm M},l}^\dagger $, with the matrices $\boldsymbol{\rho}^\alpha_l(E) $ and $\boldsymbol{\tau}_{{\rm M},l}$ respectively representing the leads electron or hole density of states  and tunnel amplitudes. 
The resulting expressions for the transmission coefficients between normal leads in terms of the Green functions are
\begin{align}
{\cal T}_{ll'}^{\alpha \beta}(E) ={}& \Tr \left[ 
\boldsymbol{\Gamma}_{l}^\alpha(E) 
{\bf G}^{r,\alpha\beta}_{\rm M}(E) 
\boldsymbol{\Gamma}_{l'}^{\beta}(E)
{\bf G}^{a,\beta\alpha}_{\rm M}(E) 
\right] 
, \, \l\neq l' 
, \nonumber\\
{\cal T}_{ll}^{\alpha \overline{\alpha}}(E) ={}&\Tr \left[ 
\boldsymbol{\Gamma}_{l}^\alpha(E) 
{\bf G}^{r,\alpha\overline{\alpha}}_{\rm M}(E) 
\boldsymbol{\Gamma}_{l}^{\overline{\alpha}}(E)
{\bf G}^{a,\overline{\alpha}\alpha}_{\rm M}(E) 
\right] ,
\end{align}
where $\alpha,\beta=e,h$ and $\overline{\alpha}=h,e$ when $\alpha=e,h$. 
The first line represents the transmission probability between different leads, while the second one corresponds to Andreev reflection processes. 
Equation (\ref{eq:currdensE}) is recovered by using the property $\sum_\beta{\cal T}_{l l}^{\alpha \beta}(E)= {\cal N}_{l}^\alpha$. 
It is important to note that Green function approaches could be very convenient to investigate interaction effects. Typically, this is done following diagrammatic methods~\cite{Merino2024}. However, discussions using that formalism are typically less transparent than the scattering approach, which instead allows us to visualize the physical processes more clearly. Finally, for linear properties, which we will mainly consider in this review, the two approaches are fully equivalent. 

One can show that in the linear regime with infinitesimal biases $V_i\to 0$ and temperature differences $\delta T_{ij}=T_i-T_j\to 0$, the heat currents satisfy $\sum_l J_{l^{}}=0$. Instead, charge conservation for charge currents is guaranteed after including currents into the superconducting leads $I_S$, i.e., $\sum_l I_{l^{}}=-I_S$. 
This is a consequence of Andreev processes changing the number of Cooper pairs in the condensate, which induces a net current in the superconducting terminals. Particle-hole symmetry of the BdG equations, see Eq.~\eqref{eq:BdG}, then yields that the scattering formalism conserves the total number of electron and hole excitations in the system.

\subsection{Linear response}
\label{sec:linear}
To discuss the charge, heat, and thermoelectrical transport, it is frequently convenient to compute the linear response coefficients. 
In the linear regime of the normal terminal biases $\delta V_{l}=(\mu_l-\mu_S)/e$ and small temperature gradients $\delta T_{ll'}\ll T$, with $T$ the operating temperature, we can write the Onsager equations~\cite{Jacquod2012Oct,Benenti2017Jun,Pierattelli2025Jun}
\begin{equation}\label{eq:onsager}
        \begin{pmatrix}
             I_l \\
             J_l
        \end{pmatrix} =
        \sum_{l'}
        \mathbf{L}_{ll'} \begin{pmatrix}
            \delta V_{l'}/T   \\
            \delta T_{ll'}/T^2
        \end{pmatrix} ,
\end{equation}
where the Onsager coefficients $\mathbf{L}_{ll'}$ are expressed in terms of the multiterminal affinities $\delta V_{l}/T$ and $\delta T_{ll'}/T^2$. 
The linear coefficients reduce to
\begin{equation}
        \mathbf{L}_{ll'}
    =T \int \frac{\md E}{2h}
        \left[
            - \frac{\partial f(E)}{\partial E}
        \right]
        \begin{pmatrix}
            e^2 \ell^+_{ll'}(E) 
            &
            eE\ell^+_{ll'}(E) \\
            eE\ell^-_{ll'}(E) 
            &
            E^2\ell^-_{ll'}(E) \\
        \end{pmatrix},
        \label{eq:conductmatr}
\end{equation}
with the integrand weighted by the derivative of the Fermi function energy (which is even in energy) and the transmission coefficients collected as~\cite{Pierattelli2025Jun}
\begin{equation}
    \ell^{\pm}_{ll'}(E) =
        N_l\delta_{ll'} - 
        {\cal T}^{ee}_{ll'}(E)
          \pm  {\cal T}^{he}_{ll'}(E) .
    \label{eq:ellpm}
\end{equation}
The sign change on ${\cal T}^{he}_{ll'}(E)$ is because holes contribute with an opposite sign to the charge but have the same contribution as electrons to the heat or energy flow. 
The Onsager coefficients can be directly connected to the transport coefficients and the most common thermoelectrical figure of merit as reported in Ref.~\onlinecite{Benenti2017Jun}.
For example, for energy independent coefficients, i.e., $\ell^{\pm}_{ll'}(E)\equiv \ell^{\pm}_{ll'} $,  the multiterminal charge conductance becomes $L^{IV}_{ll'}=G_0 T\ell_{ll'}^+$ where $G_0=e^2/h$ is the conductance quantum. Analogously, the thermal term becomes $L^{JT}_{ll'}=(\pi^2/3) (k_B^2/e^2)G_0 T \ell^-_{ll'}$. 
In the absence of superconducting leads and using the Sommerfield expansion in the low temperature limit, the multiterminal Wiedemann-Franz law $L^{JT}_{ll'}/L^{IV}_{ll'}=(\pi^2/3) (k_B^2/e^2)T$ is recovered.
However, this law is violated when superconductors are present. In fact, when Andreev contributions are present, one generally finds that $\ell^-_{ik}/\ell^+_{ik}\neq 1$. 

Since the derivative of the Fermi function is an even function in energy, comparing Eqs.~\eqref{eq:conductmatr} and \eqref{eq:ellpm} it is easy to conclude that the only finite contributions to the diagonal (off-diagonal) linear coefficients $\mathbf{L}_{ll'}$ come from the even (odd) powers of the energy of the $\ell^\pm_{ll'}$ functions. 
Consequently, when the transmission probabilities satisfy the energy inversion symmetry (EIS)
\begin{equation}
\label{eq:SymE}
{\cal T}^{\alpha\beta}_{ll'}(E)={\cal T}^{\alpha\beta}_{ll'}(-E) ,
\end{equation}
the linear thermoelectrical coefficients $L^{IT}$ and $L^{JV}$ are necessarily zero. 
Equation~\eqref{eq:SymE} represents a symmetry between positive and negative energy states. It is, however, different from the particle-hole symmetry ${\cal T}^{\alpha\beta}_{ll'}(E)={\cal T}^{-\alpha-\beta}_{ll'}(-E)$ that stems from the Nambu representation of the BdG Hamiltonian, cf. Eq.~\eqref{eq:BdG}. While particle-hole symmetry is always true for any transmission probabilities, the symmetry in Eq.~\eqref{eq:SymE} is not generic and can be broken, e.g., in multiterminal configurations or by additional degrees of freedom (e.g., spin, sublattice), enabling finite thermoelectrical coefficients. Examples of broken EIS are shown below.
However, it is important to note that the direct connection between thermoelectric effects and breaking the symmetry in Eq.~\eqref{eq:SymE} applies only in the linear regime. In the nonlinear regime or in the presence of interactions it is not easy to find general implications unless following a perturbative approach~\cite{Hwang2015Mar}. 

For the sake of a simplified analysis, we now discuss the main scattering processes in two limiting regimes, one where quasiparticle tunneling dominates and another where Andreev processes play a major role. 

\subsection{Quasiparticle tunneling}
The main contribution of the superconducting terminals tunnel coupled with normal metals is encoded in the transmission probabilities ${\cal T}^{ee}(E)$. Those quantities, for a tunneling process with a superconducting interface, would present a gapped nature with a combined peaked density of states
\begin{equation}
\label{eq:dynes}
\nu_{S}(E) = \nu_N\left|
    {\rm Re}\,\frac{E+i\Gamma_D}{\sqrt{(E+i\Gamma_D)^2-\Delta^2}}
    \right|.
\end{equation}
where $\nu_N$ is the normal density of states and $\Gamma_D$ is the small phenomenological Dynes parameter describing the quasiparticle states in the 
subgap~\cite{Dynes1978Nov}, as represented in Fig.~\ref{fig:processes}(b). They introduce strong energy filtering effects, which may be crucial to eventually determine a strong thermoelectric effect. 
Indeed, the tunnel junctions have played a crucial role in the first demonstrations of thermoelectrical effects in superconducting devices~\cite{Ozaeta2014Feb,Kolenda2016Mar}. 

Historically, tunnel junctions were shown to generate small thermoelectric effects in connection to charge imbalance effects~\cite{Thinkham1972May,Heidel1981Aug,Mamin1984Apr}.  However, the main research was initially focused on the nonlinear cooling mechanisms for  
NIS~\cite{Nahum1994Dec,Leivo1996Apr,Clark2005Apr} and SINIS configurations~\cite{Pekola2004Feb}. At the same time, the unique capability to actively control heat fluxes in superconducting tunnel junctions led to the discovery of novel heat rectification properties in NIS~\cite{Giazotto2013Dec,Martinez-Perez2015Apr} that could even be controlled by phase modulation~\cite{Strambini2014Aug,Goury2019Aug}. The experimental combination of such effects with semiconducting materials further expands the applicative possibilities in thermometry and refrigeration~\cite{Mastomaki2017Oct,Battisti2024Nov}. Those advancements show the possibility to successfully apply Peltier cooling to advanced quantum technologies such as QED resonators and, more generally, quantum computing platforms~\cite{Tan2017May,Silveri2017Sep}.
In the tunneling regime, with at least one superconducting terminal and one gapped terminal, a bipolar thermoelectrical effect induced by spontaneous particle-hole symmetry breaking has been recently predicted~\cite{Marchegiani2020Mar,Marchegiani2020Jun,Marchegiani2020Oct} and experimentally demonstrated~\cite{Germanese2022Oct,Germanese2023Jan}. 
Even if this intriguing physics depends on interaction effects \cite{Bernazzani2023Apr,Hijano2023Jun,Battisti2024Jan} it is essentially due to a nonlinear temperature difference \cite{Germanese2021Nov,Guarcello2023Oct,Antola2024Oct}. More recently, discussions on the thermoelectric properties for the junctions beyond the tunneling (opaque) barrier limit~\cite{Pershoguba2019Apr,Mukhopadhyay2022Aug} and including thermal models accounting for nonlinearities~\cite{Lucchesi2025Aug} have also been reported. 

\subsection{Andreev reflection}
\label{sec:andreev}
At low subgap energies, the Andreev reflections play a crucial role on the transport properties, with a key impact on the transmissions ${\cal T}_{l,l'}$. Indeed, Andreev reflections can also result in different contributions, depending on the system geometry. Internal reflections in SNS structures give rise to Andreev bound states (ABS) in the normal part~\cite{Kulik1969} which characterize the dissipationless transport~\cite{Beenakker1991Jun}. In the opposite sandwich, NSN, CAR gives rise to Cooper pair splitting, a sought-after source of entanglement~\cite{Recher2001Apr,Lesovik2001Dec,Prada2004Aug}, for which EC contributions need to be suppressed. 

The direct contribution of LAR processes to thermoelectric phenomena appears unlikely, as the inherent superposition of quasiparticle and quasi-hole states in these processes enforces an EIS, as described by Eq.~\eqref{eq:SymE}. Nevertheless, Andreev reflections play a crucial role in enabling phase-dependent responses and other nonlocal effects, as will be discussed in the following sections. 

Unlike conventional even-frequency superconductivity, odd-frequency pairing involves a pair amplitude that is antisymmetric under time (or frequency) exchange, allowing spin-triplet $s$-wave states compatible with Fermi statistics. 
Breaking spin-rotation symmetry at a superconducting junction induces spin-triplet Cooper pairs that enable Andreev processes with a significant thermoelectric contribution~\cite{Hwang2018Oct, Savander2020Dec}. In such scenarios, a finite thermopower arising from Andreev-like processes is only possible in the presence of odd-frequency pairing~\cite{Hwang2018Oct}. 
This principle holds generally for superconducting systems. Importantly, a finite Andreev thermopower requires coexistence of even- and odd-frequency components, a typical feature of proximity-induced unconventional pairing. 

Finally, in superconducting heterostructures the Andreev approximation assumes that the chemical potential is much larger than the energies involved ($\mu\gg E,\Delta$). This means that electrons and holes in Andreev reflections have nearly the same momentum and follow the same path—called \textit{retro-reflection}. In Sec.~\ref{sec:2degs}, we examine what happens in Dirac materials, where this approximation breaks down because of their linear energy-momentum relation.

\section{Quantum dots and single-electron transistors}
\label{sec:QD}
A simple way to break the EIS in the absence of an applied voltage is by the spectral properties of a nanostructure placed in the normal-superconductor interface. The simplest case is the discrete level of a quantum dot~\cite{Kouwenhoven1997,Ihn2009,Martin-Rodero2011Dec} in the Coulomb blockade regime. Usually, for simplicity, the quantum dot states are limited to host up to two electrons since the Coulomb properties repeat periodically for any even number of electrons~\cite{Kouwenhoven1997}. The strong energy filtering of quantum dots has been demonstrated to achieve highly efficient thermoelectric capabilities when the level is tuned in the vicinity (at energies of a few $\kBT$) of the chemical potential~\cite{Staring1993Apr,Dzurak1993Sep,Josefsson2018Oct} in semiconductors. The same effect can, in principle, be exported to tunnel-coupled N-QD-S interfaces; however, the transport of quasiparticles will be strongly suppressed by the gap (unless the temperature of the normal electrode is comparable to $\Delta$).

\begin{figure}[t]
  \includegraphics[width=\linewidth]{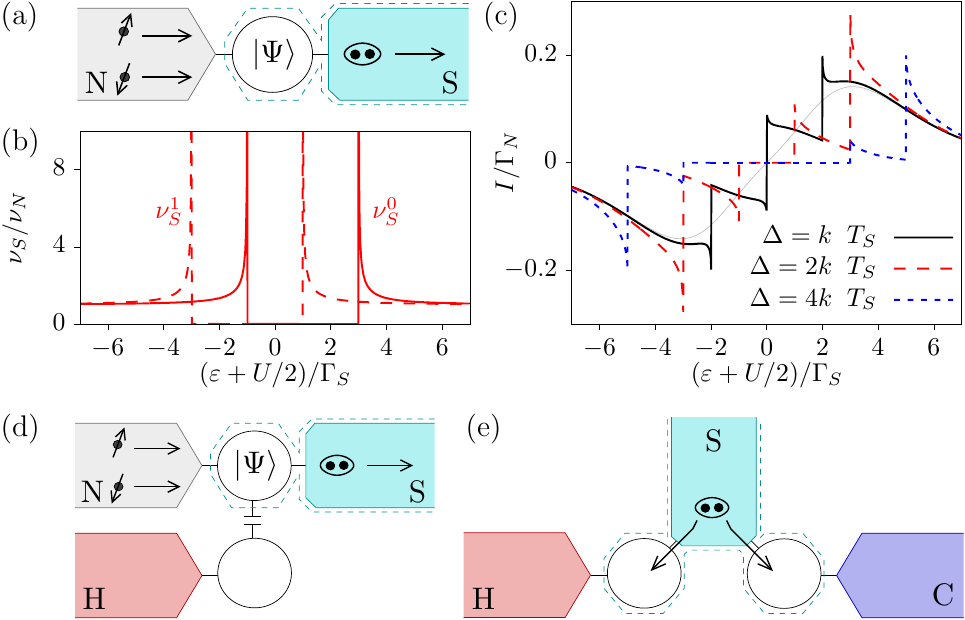}
  \caption{(a) Scheme of a two-terminal thermoelectric engine with a proximized quantum dot. (b) Superconducting density of states for $\Delta=2\kBT_S$, and (c) generated current for different gaps in the quasiparticle regime, for $U=2\kBT_S$ and $T_N=2T_S$. The gray line in (c), corresponding to the {\it all-normal} case $\nu_S=\nu_N$, is shown for reference. The tunneling rate through the normal contact is $\Gamma_N=2\pi|\tau_N|^2\nu_N/\hbar$. (d) Nonlocal three-terminal heat engine generating current between terminals $N$ and $S$ from the heat injected from $H$ via the capacitive coupling of two quantum dots. (e) Cooper pair splitter serving as a nonlocal thermoelectric and refrigerator. }
  \label{fig:proxQD}
\end{figure}

Increasing the coupling to the superconductor, the quantum dot states get hybridized due to the proximity effect, allowing for the exchange of Cooper pairs, as illustrated in Fig.~\ref{fig:proxQD}(a). When the gap is the largest energy scale $\Delta\to\infty$, this process is described by an effective model that replaces the 
Hamiltonian of the superconductor by the pairing term~\cite{Rozhkov2000Sep,Braggio2011Jan,Rajabi2013Aug,Nigg2015Mar}
\begin{equation}
    \Gamma_{S}(\hat{d}_{\uparrow}^{\dagger}\hat{d}_{\downarrow}^{\dagger}+{\rm {h.c.})},
\end{equation}
where $\hat{d}_{\sigma}^{ }$ annihilates an electron of spin $\sigma$ in the quantum dot and $\Gamma_S$ is the coupling. 
Note that this effective model describes the coherent coupling of the quantum dot states with the superconducting condensate, which acts as a coherent source of Cooper pairs.
Adding it to the 
Hamiltonian of the quantum dot
\begin{equation}
\label{eq_qd}
\hat{H}_{{\rm qd}}=\sum_\sigma\varepsilon_\sigma\hat{n}_\sigma+U\hat{n}_{\uparrow}\hat{n}_{\downarrow},
\end{equation}
with the energy of the level $\varepsilon_\sigma$, the intradot Coulomb interaction $U$, and the number operators $\hat{n}_\sigma=\hat{d}_{\sigma}^{\dagger}\hat{d}_{\sigma}^{ }$, we get the Hamiltonian of the proximized quantum dot. 
The interaction term lifts the single particle description of the scattering theory, so many body descriptions are required e.g., Keldysh nonequilibrium Green functions~\cite{Merino2024} or master equations~\cite{Blum2012}.
This Hamiltonian  can be diagonalized, giving the even-parity ABS 
\begin{align}
    \left|\pm\right\rangle  =\mathcal{N}_{\pm}^{-1}(A_{\pm}\left|0\right\rangle -\Gamma_{S}\left|2\right\rangle ),
    \label{eq:evenst}
\end{align}
where $\mathcal{N}_{\pm}$ is imposed by the normalization of the $\left|\pm\right\rangle$ states, 
with energies 
\begin{equation}
E_{\pm}=A_{\pm}\equiv\tilde{\varepsilon}\pm\sqrt{\tilde{\varepsilon}^{2}+\Gamma_{S}^{2}},
\end{equation}
where $\tilde{\varepsilon}=(\sum_\sigma\varepsilon_{\sigma}+U)/2$. These superpositions of states with 0 and 2 electrons mediate the exchange of Cooper pairs with the superconductor which can be controlled by gate voltages applied to the quantum dot. The odd states $|\sigma\rangle$, with eigenenergies $E_{\sigma}=\varepsilon_{\sigma}$, are not coupled. 
Note that the property $A_+\to-A_-$ by reversing $\tilde\varepsilon\to-\tilde\varepsilon$, which can be tuned by the quantum dot gate voltages, defines the particle-hole symmetry point. The consequence of this symmetry is that the energy spectrum of the ABS is symmetric in energy. So even if the transport properties are clearly affected by the proximity effect~\cite{Fazio1998Mar,Braggio2011Jan} this even symmetry of the spectrum may determine the EIS of Eq.~\eqref{eq:SymE}, suppressing the linear thermoelectrical effects. However, it is important to note that in the nonlinear regime thermoelectricity can emerge~\cite{Hwang2015Mar,Panu2024Oct}.

\subsection{Breaking energy symmetry in Coulomb blockaded quantum dots}
\label{sec:QD_CB}
In this section we discuss, even in the presence of superconducting proximity effects, how to break EIS. To do that it is important to include in the description the tunneling through the contacts to normal reservoirs $l$ which corresponds to a transition process between even and odd parity (number of electrons) states. The tunneling term is 
\begin{equation}
\hat{H}_{tun}=\sum_{lk\sigma}\tau_{l}\hat{d}_\sigma^\dagger\hat{c}_{k\sigma}+h.c.,
\end{equation}
where $\hat{c}_{k\sigma}$ annihilates an electron of wavenumber $k$ and spin $\sigma$ in the reservoir. In the sequential tunneling limit at weak coupling, the transition rates for the transition $|\kappa\rangle\to|\lambda\rangle$ due to an electron or a hole entering/outgoing the quantum dot from the reservoir are given by Fermi's golden rule~\cite{Hussein2016Dec}
\begin{align}
\label{eq:gammaNe}
\gamma_{\lambda\kappa}^{l\alpha}&=\frac{2\pi}{\hbar}|\tau_{l}\langle\lambda|\hat{\delta}_{\sigma}^{\alpha}|\kappa\rangle|^{2}\nu_{l}f_{l}^\alpha\left(E_{\lambda}-E_{\kappa}\right)
\end{align}
where $\hat\delta_{\sigma}^e=\hat{d}_{\sigma}^{\dagger}$, $\hat\delta_{\sigma}^h=\hat{d}_{\sigma}^{}$ and $f_{l}^\alpha(E)$
is the Fermi function for the normal reservoir with density of states $\nu_{l}$ at electrochemical potential $\mu_{l}$ and temperature $T_{l}$. 
Clearly, $|\langle\sigma|\hat{d}_{\sigma}^\dagger|\pm\rangle|^2\propto{A}_{\pm}^2$ is different from $|\langle\sigma|\hat{d}_{\sigma}|\pm\rangle|^2\propto\Gamma_S^2$, except for the electron-hole symmetry point $\tilde\varepsilon=0$. Hence, the electron/hole contribution of each transition can be controlled by tuning the quantum dot gate voltage. This will also affect the occupation of the different states. 
In the sequential tunneling regime, one uses these rates to write a master equation $\dot\rho_{\kappa\kappa}=\sum_{l\alpha\lambda}(\gamma_{\kappa\lambda}^{l\alpha}\rho_{\lambda\lambda}-\gamma_{\lambda\kappa}^{l\alpha}\rho_{\kappa\kappa})$ for the state occupations, $\rho_{\kappa\kappa}$~\cite{Blum2012}, and writes the particle and heat currents out of terminal $l$~\cite{Schaller2014Jan}
\begin{align}
\label{eq_I}
    I_l&=\underset{\lambda,\kappa}{\sum}\left(\gamma_{\lambda\kappa}^{l e}-\gamma_{\lambda\kappa}^{l h}\right)\rho_{\kappa\kappa},\\
\label{eq_J}
    J_{l} &=\underset{\lambda,\kappa}{\sum}\left(E_{\lambda}-E_{\kappa}-\mu_l\right)\left(\gamma_{\kappa\lambda}^{le}\rho_{\lambda\lambda}-\gamma_{\lambda\kappa}^{lh}\rho_{\kappa\kappa}\right).
\end{align}
using the steady state solution, $\dot\rho=0$.

Consider the two-terminal setup of Fig.~\ref{fig:proxQD}(a). The thermoelectric contribution of Cooper pairs at low energies is expected to be suppressed for weak interactions~\cite{Hwang2015Mar} due to the intrinsic electron-hole symmetry of proximized states. When electron-electron interactions are strong, the onset of Coulomb blockade introduces the presence of localized magnetic states~\cite{Anderson1961Oct} and the single-particle picture of scattering theory of Sec.~\ref{sec:physics} does not hold. The presence of such correlations may enable the lifting of the the EIS. This possibility has however attracted little attention so far.

The regime dominated by quasiparticles at higher temperatures or weak coupling to the superconductor offers a clearer picture. There, ignoring the Andreev reflection contributions, the particle tunneling to the superconductor is governed by the BCS density of states of Eq.~\eqref{eq:dynes}. 
The tunneling rates will be proportional to $\nu_S^n=\nu_S(\varepsilon+nU)$, where $n$ is the minimum occupation of the quantum dot before or after the process, see Fig.~\ref{fig:proxQD}(b). The effect of the gap in $\nu_S^n$ prevents charge from flowing for quantum dot energies around the chemical potential except when $\Delta\leq U/2$. In a sequential tunneling description, it reads $I\propto\sum_{nn'}(f_{Nn}+\nu_S^nf_{Sn})\nu_{S}^{n'}(f_{Nn'}-f_{Sn'})(1-\delta_{nn'})$, where $f_{ln}=f_l(\varepsilon+nU)$, see Fig.~\ref{fig:proxQD}(c). 
The current is hole(electron)-like for $\tilde\varepsilon<0(>0)$, as would correspond to an all-normal configuration~\cite{Beenakker1992Oct}, with sharp features corresponding to the peaked density of states at the gap borders.

In realistic configurations, both quasiparticle and Cooper pair processes coexist, requiring more involved treatments~\cite{Verma2022Feb, Verma2024Mar}. The nonlinear thermoelectric response has been investigated in single quantum dots with Coulomb correlations working as power generators~\cite{Krawiec2008,Hwang2015Mar,Verma2022Feb}, refrigerators~\cite{Hwang2023Jun,Verma2024Mar} and thermoelectric diodes~\cite{Hwang2016Sep}. More complicated quantum dot arrangements have also been proposed to enhance the thermoelectric response~\cite{Yao2018Nov}. 
Ring structures can also be used for further control of the device via the Aharonov-Bohm effect~\cite{Yao2018Apr,Blasi2022Dec,Yao2023}.

Multiterminal configurations have also been considered. In particular, three-terminals~\cite{Gramich2017Nov} allow one to define local and nonlocal (when the response is measured in terminals different from the heat source) Seebeck responses~\cite{Michalek2016Jun} or couple to spin degrees of freedom when one of them is ferromagnetic~\cite{Wysokinski2012Jul,Sonar2024Mar}.
One can also use the LAR to separate heat and charge flows: heat injected from one terminal is Andreev-reflected in the superconductor contact, with the other terminal working as a voltage probe~\cite{Mazza2015Jun}. Coulomb blockade also controls the heat flow in metallic islands coupled to a phonon bath~\cite{Goury2019Aug} when using superconducting quantum interference single-electron transistors~\cite{Enrico2016Jun,Enrico2017Oct} to tune the superconducting gaps in proximized weak links.

When the superconducting terminals are more than one it is als possible to establish phase differences between superconducting terminals. In such case it is possible to investigate phase-dependent effects on heat and charge transport and thermoelectrics~\cite{Kamp2019Jan,Bauer2019Jun,Bauer2021Nov,Kumar2025Jan}. Intriguing thermoelectrical effects are also possible when superconductors and quantum dots are coupled with ferromagnetic materials~\cite{Trocha2017Apr,Hwang2023Jun,Wojcik2014Apr}. Nonlocal cooling mechanism in ballistic Andreev interferometers have also been proposed recently~\cite{Cioni2025Jul}. 

\subsection{Interacting nonlocal heat engines}
\label{sec:QD_interact}
Three-terminal setups also allow to separate the system into a heat source that injects heat but no charge current, and the electronic circuit, which can this way be isothermal. This is the geometry of a thermocouple~\cite{Sothmann2014Dec,Balduque2025Apr}, realized experimentally in semiconductor samples~\cite{Sanchez2011Feb,Thierschmann2015Oct}. Electrical isolation of the heat source is achieved by using capacitively coupled quantum dots: one of them (the passive) supports a charge current between the $N$ and $S$ terminals, with the other (the active) being connected to charge fluctuations from the hot terminal, $H$, as sketched in Fig.~\ref{fig:proxQD}(d). The transfer of heat is then mediated by the Coulomb interaction between electrons the different quantum dots, $U_{ap}$. A finite zero-bias current will flow under the appropriate symmetry breaking, namely that the rate for particle transfers between $N$ and $S$ after exchanging an energy $U_{ab}$ with $H$ is different for electrons and holes~\cite{Tabatabaei2022Sep}.

The working mechanism is different depending on whether the injected current is due to quasiparticles or to Cooper pairs. In the former case, the presence of the gap acts as an energy filter which is sufficient to have a net quasiparticle current~\cite{Bhandari2018Jul}, and the absorption of $U_{ap}$ provides the necessary energy to overcome it when $\varepsilon_a+U_{ap}\sim\Delta$. This mechanism introduces a more robust way to control the asymmetry as compared to all-semiconductor versions~\cite{Thierschmann2015Oct} that improves the thermoelectric performance in terms of both the generated power and the efficiency. 
The Cooper pair transport dominates at low temperatures and energies close to $\mu_S$. It can hence be understood in the infinite gap regime, in terms of the effect that the charge fluctuations in the active dot have on the proximized states in the passive quantum dot. A charge in the active dot shifts the energy on the active one by $U_{ap}$, which furthermore affects the weights in the even superpositions. This way the contribution of electrons and holes to each tunneling event are modified and eventually lift the symmetry between positive and negative energy processes. This current is hence a consequence of the interplay of quantum superpositions and nonequilibrium fluctuations. Remarkably, the quasiparticle and Cooper pair contributions have opposite signs, so the current serves as a measure of their relative dominance~\cite{Tabatabaei2022Sep}. 

The system can be used in reverse as a nonlocal Peltier refrigerator that cools a metallic island capacitively coupled to the charge conductor~\cite{Sanchez2017Nov,Bhandari2018Jul}. Being a multiterminal conductor, it is able to perform as an absorption refrigerator or a heat pump, or even in {\it hybrid} operations~\cite{Manzano2020Dec} with pairs of these useful tasks obtained simultaneously~\cite{Tabatabaei2022Sep,Lopez2023Jan}. The same mechanism can be used to define an autonomous (quasiparticle) Maxwell demon that generates power between two superconductors without them absorbing energy from a pair of capacitively coupled quantum dots supporting a heat flow~\cite{Sanchez2019Oct}.

\subsection{Thermoelectricity by Cooper pair splitting}
\label{sec:cps}
The superconductor can act as a mediator for particle and heat nonlocal transport between two normal reservoirs. This is due to two characteristic processes: CAR and EC. The former one, due to the splitting of Cooper pairs~\cite{Recher2001Apr,Lesovik2001Dec}, has attracted a big interest for being a possible source of entangled electrons~\cite{Recher2001Apr,Lesovik2001Dec,Chtchelkatchev2002Oct,Samuelsson2003Oct,Bignon2004Jul}. To select CAR over EC and avoid the two electrons to tunnel to the same lead, one uses Coulomb blockade quantum dots~\cite{Hofstetter2009Oct,Das2012Nov,Fulop2015Nov,Deacon2015Jul,Herrmann2010Jan,Schindele2012Oct,Tan2015Mar,Ranni2021Nov,Wang2022Dec,Bordoloi2022Dec,deJong2023Oct}, see Fig.~\ref{fig:proxQD}(e).

This geometry can be used to define a nonlocal thermoelectric engine: a temperature difference between the two normal reservoirs generates a current out of the superconductor~\cite{Cao2015Nov,Sanchez2018Dec,Hussein2019Feb,Kirsanov2019Mar,An2023Apr}. Experiments have confirmed this mechanism~\cite{Tan2021Jan}. When the quantum dot level energies are antisymmetric around the superconductor electrochemical potential, energy-conserving Cooper pair splitting processes are privileged over EC, such that the same current flows ingoing (outgoing) from (to) the two normal leads into (from) the superconducting terminal~\cite{Sanchez2018Dec,Hussein2019Feb}. This nonlocal splitting explicitly breaks the EIS such that, e.g., ${\cal T}_{HC}^{eh}(E)\ne{\cal T}_{HC}^{eh}(-E)$. Heat is hence converted into positively cross-correlated charge flows~\cite{Golubev2023Jan} and can also lead to coherent Peltier cooling, which serves as a prove of Cooper pair splitting~\cite{Sanchez2018Dec}. Differently EC contributes to the local thermoelectric response between the normal reservoirs~\cite{Sanchez2018Dec,Hussein2019Feb,Kirsanov2019Mar}, in a similar way to what one expects from all-semiconductor double quantum dots~\cite{Thierschmann2013Dec,Dorsch2021Jan}. Configurations with EC and CAR having similar contributions can be beneficial for enhancing power and efficiency when the hot terminal is a voltage probe~\cite{Sanchez2018Dec} or determining strong nonlocal responses from heating effects~\cite{Kolenda2013Nov}.

\subsection{Detection of Majoranas}

Since the early investigations into Majorana states in condensed matter systems, thermoelectric effects have been identified as a potential signature of their presence.
Majorana signatures have been pointed out in thermal conductance \cite{Fu2008Mar,Bauer2021Nov}, voltage thermopower \cite{Hou2013Aug,Sela2019Oct}, or the violation of the Wiedemann-Franz law \cite{Buccheri2022Feb,Benjamin2024Apr}.
In quantum-dot-based multiterminal setups, it has been suggested that a key signature for detecting Majorana zero modes (MZMs) is the opposite sign behavior of the Seebeck coefficient as a function of the quantum dot energy level, in contrast to what is observed in conventional superconductors \cite{Leijnse2014Jan,Lopez2014May,Valentini2015Jan,Ricco2018Feb,Sun2021Jun,Ri2019May,
Smirnov2021Feb,Grosu2023Jan}. Most of these works just consider the effect of the MZMs and neglect the contribution of quasiparticle states above the gap, whose effect in the out of equilibrium response of these systems has remained largely unexplored \cite{Trocha2024Apr,Trocha2025Jan} but can be included by means of Green function techniques \cite{Zazunov2016Jul}. The thermoelectric response of more complex geometries including a double QD connected in series \cite{Majek2022Feb} or in parallel \cite{Klees2024Aug} with a topological superconductor hosting MZMs has also been explored. 

\section{1D Electron systems}
\label{sec:wires}
Electron systems constrained in 1D are unique platforms to investigate how dimensional reduction will crucially affect transport. Early experiments of conductance quantization in quantum point contacts~\cite{vanHouten1992Mar} and the development of edge state physics in quantum Hall samples and in 2D topological insulators, postulated 1D systems as controllable cases of quantum transport. More recently, the possibility to localize Majorana bound states (see before) at the boundary of 1D helical channels in the presence of magnetic fields and superconducting proximity further attracted attention over the thermoelectric properties of these systems also in the presence of superconductors~\cite{Marciani2014Jul,Lopez2014May,Leijnse2014Jan,Valentini2015Jan,Chi2020Dec,Ramos-Andrade2016Oct,Majek2022Feb,Buccheri2022Feb,Smirnov2023Apr,Wang2019May,Benjamin2024Apr,Chi2024Jun,Klees2024Aug}.

\subsection{Helical edge states}
Chiral and helical edge states of two-dimensional systems in the quantum Hall and the spin quantum Hall regime 
define the paradigm for one-dimensional ballistic transport \cite{Buttiker1988Nov}.  The quantum spin Hall effect
is realized in two-dimensional (2D) topological insulators (2DTI) and, typically, the fundamental ingredient is
the spin-orbit coupling \cite{Kane2005Nov,Kane2005Sep,Bernevig2006Dec,Konig2007Nov}. This topological phase is
due to the existence of Kramers pairs of helical edge modes, which can be described in terms of a 1D Dirac Hamiltonian density of the form
\begin{equation}
H_0(x) = v_F  \left(-i \hbar \partial_x \right)\sigma^z,
\end{equation}
where the Pauli matrix acts on the two components of the spinors $\psi^T=\left(\psi_\uparrow, \psi_\downarrow \right)$, and 
is parallel to the natural quantization axis of the 2DTI (usually defined by the spin-orbit coupling).

\begin{figure}  
\includegraphics[width=\linewidth]{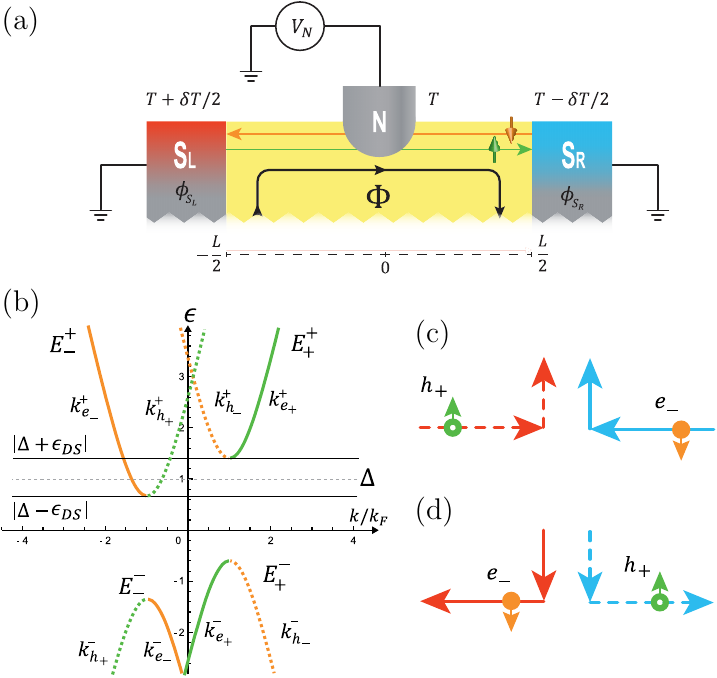}
  \caption{Sketch of the setup considered in Refs. \cite{Blasi2020Jun,Blasi2021Jun}. 
A Kramers pair of helical edge states of the quantum spin Hall effect is contacted by two superconductors at different temperatures, $T_{S_L, S_R}= T \pm  \delta T/2$ and with a normal-metal probe at temperature $T_N=T$ at which a bias voltage $V_N$ is eventually applied/generated. The structure is threaded by a magnetic flux, which induces a Doppler shift in the edge states in addition to a Josephson phase difference applied between the two superconductors. (b) Dispersion curves for quasiparticles $e_\pm$ (solid lines) and quasiholes $h_\pm$ (dashed lines) within the proximized spin-Hall region.
Transport processes are depicted in (c) for $V_N = 0$; $\delta T \neq 0$ and in (d) for $V_N \neq 0$; $\delta T = 0$, when the spectrum for $e_+; h_-$ is assumed fully gapped. With courtesy of Ref.~\onlinecite{Blasi2020Jun}.
 }
  \label{fig:s-ti-s}
\end{figure}

In Ref.~\onlinecite{Gresta2019Oct}, it was shown that a two-terminal device of a 2DTI hosting a magnetic island is an
optimal thermoelectric heat engine. Here, we briefly review the thermoelectric properties of 
the device studied in Refs. ~\cite{Blasi2020Jun,Blasi2021Jun}, where the 2DTI is contacted to superconductors forming a Josephson junction, cf. Fig.~\ref{fig:s-ti-s}(a).
The edge modes are contacted by left ($S_L$) and right ($S_R$) $s$-wave type superconductors. Considering the Nambu basis $\psi^T=\left(\psi_\uparrow, \psi_\downarrow, \psi^{\dagger}_\downarrow, -\psi^{\dagger}_\uparrow \right)$, the BdG Hamiltonian now becomes 
\begin{equation}
    H_\text{BdG}(x) = \left[H_0(x,p_{DS})\tau^z + \Delta^\prime(x)\tau^x +\Delta^{\prime \prime}(x) \tau^y\right]/2,
\end{equation}
where $\Delta(x)=\Delta\left[\theta(-x-L/2)e^{i\phi_{L}}+ \theta(x-L/2)e^{i\phi_{R}}\right]$ is the pairing potential induced on the edge modes within the region where they are proximized to the superconducting contacts. A phase bias $\phi=\phi_{L}-\phi_{R}$ is considered between the two superconductors, which define real and
imaginary parts of the pairing potential ($\Delta^\prime(x)$ and $\Delta^{\prime \prime}(x)$, respectively). The other ingredient is a {\em Doppler shift}~\cite{Tkachov2015Jul} $p_{DS}=(\pi\hbar/L)\Phi/\Phi_0$ generated by the magnetic flux $\Phi$ within the Josephson junction region of TI between the superconductors, i.e., $-L/2\leq x\leq L/2$, with $\Phi_0$ the magnetic flux quantum. 
This Hamiltonian is particle-hole symmetric, and the naive expectation for the symmetric right-left configuration is the absence of any thermoelectric local response between the superconducting terminals when a temperature bias is imposed at the superconductors. However, in Refs.~\cite{Blasi2020Jun,Blasi2021Jun}, it was shown that nonlocal thermoelectricity is still possible in such a system by adding a third normal-metal terminal, see Fig.~\ref{fig:s-ti-s}(a). 
The mechanism generating the nonlocal thermoelectric response is understood by analyzing the spectrum of the
full Hamiltonian, shown in Fig.~\ref{fig:s-ti-s}(b). 
We can identify the particle-hole symmetry of 
the BdG Hamiltonian, with positive and negative energies related by  $\varepsilon_k=-\varepsilon_{-k}$.
In the different branches of the spectrum, the electron-like (e) and hole-like (h) character of the quasiparticles is indicated with solid and dashed lines, respectively. 
We can identify two cones. Each one has left ($-$) and right ($+$) moving quasiparticles (plotted in different colors). The Doppler shift breaks the symmetry under the inversion $k \leftrightarrow -k$ of the spectrum. As a consequence, the main contribution to the transport window originates in a single cone hosting states of quasielectrons and quasiholes traveling in opposite directions. Importantly for our purposes, however, the two types of particles are injected in different reservoirs at different temperatures. Since they come from opposite thermal sources they introduce an extra sign when the current is computed in the normal middle terminal such as ${\cal I}_N\propto\sum_\alpha \alpha( {\cal T}^{e\alpha}_{NL}-{\cal T}^{e\alpha}_{NR})$ with respect to the local term ${\cal I}_{R}\propto\sum_\alpha \alpha ({\cal T}^{e\alpha}_{RL}+{\cal T}^{e\alpha}_{LR})$. Therefore, even if the local thermoelectric response is null a net particle charge flux is generated 
at the normal probe (nonlocal response)~\cite{Mateos2024Aug}, as sketched in Fig.~\ref{fig:s-ti-s}(c). The reciprocal process, injecting a net charge current at the normal probe, is sketched in Fig.~\ref{fig:s-ti-s}(d). Such a flux is converted into uncompensated fluxes of quasielectrons and quasiholes in the proximized TI, generating a net heat flux between the two superconducting contacts. 

The electron current in the normal probe evaluated in the linear-response regime is given by the following relations
\begin{gather}
\label{jn-lr}
\begin{aligned}
    J^0_N &=L^{(nl)}_{11}\left(V_N/T\right)+L_{12}^{(nl)}\left(\delta T /T^2\right), \\
    J^0_N &=L_{21}^{(nl)}\left(V_N/T\right)+L_{22}^{(nl)}\left(\delta T /T^2\right),
\end{aligned}
\end{gather}
where the Onsager coefficients $L_{ij}$ are generalized to include nonlocal response in this three-terminal configuration.
They depend on the Doppler shift $p_{DS}$, the phase bias $\phi$, the 
pairing amplitude $\Delta_0$, as well as the temperature $T$. Interestingly, the nonlocal off-diagonal elements are related by symmetry as follows: $L^{(nl)}_{12}(\Phi,\phi)=-L^{(nl)}_{21}(-\Phi,-\phi)$, as required by the general symmetries for linear coefficients~\cite{Jacquod2012Oct}. 
A finite nonlocal thermoelectric response is predicted for $\phi\neq 0 \; {\rm (mod)}\pi $ and/or $\Phi \neq 0 \;{\rm (mod)}\pi$ leading to a Seebeck coefficient $S=\left(L_{12}/L_{11}\right)/T$ close to 100 $\mu \text{V}/\text{K}$ at low temperatures (10\% of the critical temperature of the superconductor). Importantly, this effect is a direct consequence of the helical nature of the edge modes of the TI, which is an appealing hallmark from the experimental point of view. It has been shown that this mechanism can be employed in the non-linear regime to implement a nonlocal thermoelectric heat engine~\cite{Blasi2022Dec}.

Thermal transport, albeit not really thermoelectric, has also been studied in this type of systems~\cite{Sothmann2016Aug,Sothmann2017Feb,Bours2019Apr}.
More complex devices including the effect of magnetic 
islands in the Josephson junction generating Jackiw-Rebbi type of topological zero modes have also been predicted~\cite{Mateos2024Aug}.
In addition, the phase-dependence of charge and heat currents in thermally biased short Josephson junctions with the helical edge states of a 2DTI has also been recently analyzed~\cite{Gresta2021Feb}. 

Another related proposal is based on a thermally biased Andreev interferometer with a middle ferromagnetic island on top of the helical edge of a 2DTI~\cite{Keidel2020Apr}. 
Without the magnetic impurity, the thermoelectric conversion is analyzed in Ref.~\onlinecite{Dutta2023Aug}. 
In the presence of the magnetic domain, a spin-polarized thermal supercurrent emerges between the superconductors due to a unique interference effect of CARs, that is absent for EC processes. 
The resulting thermoelectric effect can be turned on and off by tuning the phase difference between the superconducting leads and does not rely on manipulating the ferromagnetic domain. 

\subsection{Superconducting helical wires}
\begin{figure}
  \includegraphics[width=\linewidth]{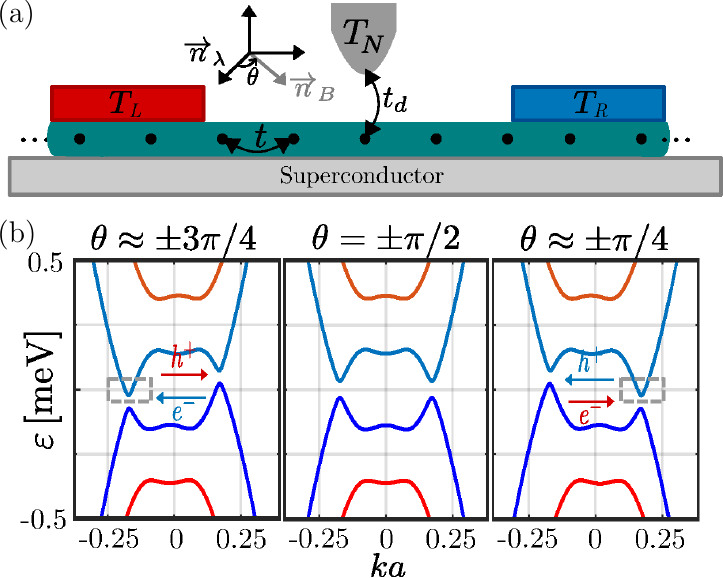}
  \caption{(a) Sketch of the setup considered in Ref.~\onlinecite{Mateos2024Aug}. 
A semiconducting wire with SOC and magnetic field is contacted to a superconductor. A temperature bias $ T \pm  \delta T/2$ is imposed and the electrical current is generated
at a normal-metal probe at temperature $T_N=T$ at which a bias voltage $V_N$ is applied.  
(b) Spectrum for different angles between the SOC and the magnetic field. The central panel corresponds to a perpendicular orientation of the magnetic field with respect to the direction of the SOC, while the other panels correspond to departures from this configuration, leading to the formation of Bogoliubov Fermi points. This situation is similar to the one illustrated in Fig. \ref{fig:s-ti-s}(b) and favors a nonlocal thermoelectric response. With courtesy of Ref.~\onlinecite{Mateos2024Aug}
 }
  \label{fig:s-bog}
\end{figure}

A similar effect to the one discussed in the previous section is predicted to take place in semiconducting wires with Rashba spin-orbit coupling (SOC) proximized to superconductors and under the effect of a magnetic field \cite{Mateos2024Aug}. 
These types of structures are used to implement the 1D topological superconducting phases hosting Majorana edge modes experimentally~\cite{Alicea2012Jun,Flensberg2021Oct},
motivated by the model proposed in Refs.~\onlinecite{Oreg2010Oct,Leijnse2014Jan}. 
Experimentally, these wires are fabricated with GaAs doped with In and proximized with a $s$-wave superconductor of gap $\Delta$ \cite{Mourik2012Apr,Deng2016Dec,Chen2017Sep,Chen2019Sep,Nichele2017Sep,Vaitiekenas2020Mar}. The Hamiltonian reads
\begin{equation}
    H_\text{BdG}(k) = \left[\left(\xi_k{-}\lambda_k \vec{n}_\lambda \cdot \vec{\sigma}\right)\tau^z - \Delta_B \vec{n}_B\cdot \sigma + \Delta
    \tau^x \right]/2,
\end{equation}
where $\lambda_k=\alpha k$ represents the SOC acting on the spin degrees of freedom along the direction $\vec{n}_\lambda$ and $\Delta_B= g\mu_B B$ is the Zeeman term associated to the magnetic field $B$ acting along the direction $\vec{n}_B$. $\xi_k=k^2/2m-\mu$ is the kinetic energy relative to the chemical potential $\mu$. When such wires are contacted at the ends with reservoirs at different temperatures $T \pm  \delta T/2$
and contacted at an intermediate point with
a normal conducting probe at temperature $T$, see Fig.~\ref{fig:s-bog}(a), a nonlocal thermoelectric response is predicted when the angle between the magnetic field and the SOC departs from perpendicular orientations, $\theta =\cos^{-1}(\vec{n}_B\cdot \vec{n}_\lambda)\neq n \pi/2$. 

The origin of the nonlocal response can be understood by analyzing the spectrum.  Without superconductivity, the spectrum contains two bands and a gap defined by the combined effect of the SOC and the Zeeman field.
Adding superconductivity and expressing the Hamiltonian in Nambu basis leads to the spectra shown in Fig.~\ref{fig:s-bog}(b).  
The relevant situation for thermoelectric response takes place when the chemical potential is close to the bottom of the lower band, precisely the case illustrated in the first and last panels of Fig.~\ref{fig:s-bog}(b). It is important to notice that when the magnetic field and the SOC are exactly perpendicular [see central panel in Fig.~\ref{fig:s-bog}(b)] the spectrum is fully symmetric with respect to changing $k \rightarrow -k$. Instead, for other angles, one of the quasiparticle cones with positive energy moves to lower energies and eventually crosses zero, defining two Bogoliubov Fermi points. This situation is identified with a gray dashed-line box in Fig.~\ref{fig:s-bog}(b) and is similar to the one discussed in the analysis of Fig.~\ref{fig:s-ti-s}. In fact, the low-energy sector of the spectrum is composed of a single branch of quasielectrons moving to the right and quasiholes moving to the left. These uncompensated fluxes generate an electron current at the normal lead described in linear response by
Eq.~(\ref{jn-lr}). Instead, for higher values of the chemical potential, the two bands of the model without superconductivity contribute to the transport processes. In this situation, there are a pair of left- and right-moving quasielectrons as well as a pair of
left- and right-moving quasihole branches. The net charge flux is compensated in this case and the electron current in the normal probe
is vanishingly small. Interestingly, the range of chemical potentials for the topological phase to take place is $\mu^2 \leq \Delta_B^2-\Delta^2$ and coincides with the one for which a non-vanishing thermoelectric response is expected. It is also interesting that the
range of angles for which the topological phase takes place is complementary to the one with important thermoelectric response. Namely, the topological phase is predicted within $| \theta-\pi/2 |\leq \delta_c$, while the thermoelectric response is predicted within
$| \theta-\pi/2 |> \delta_c$. 
Note that this nonlocal thermoelectric
effect is not a signature of Majorana fermions {\it per se}. Nevertheless, it provides very relevant information to characterize the topological phase and
the relative alignment between the SOC and the magnetic field to identify where in the wide parameter space the topological transition should localize. The nonlocal thermoelectric transport requires a proximized helical system in the electron bands, which is also a necessary
condition for the nontrivial topological phase.

A recent experimental study of Bogoliubov Fermi surfaces in two-dimensional systems \cite{Phan2022Mar} has motivated the proposal of thermoelectric effects to explore this interesting mechanism \cite{Pal2024Dec}.

\subsection{Chiral edge states}
Recently, a peculiar nonlocal thermoelectric property of edge states of a quantum Hall system (QHS) contacted to a superconducting terminal has been analyzed~\cite{Panu2024Oct}. The electrical transport of this system is a topic of increasing interest nowadays and is the focus of many recent theoretical \cite{Ma1993Mar,Manesco2022Sep,Kurilovich2023Apr,Kurilovich2023Sep,Schiller2023Apr,Tang2022Dec,Giazotto2005Aug,vanOstaay2011May,Zhang2019Jun,PeraltaGavensky2020Aug,Galambos2022Aug,Khrapai2023Jun,David2023Mar,Michelsen2023Jan,Blasi2023Aug,Arrachea2024Feb} and experimental~\cite{Wan2015Jun,Amet2016May,Park2017Sep,Lee2017Jul,Guiducci2019Jun,Zhao2020Aug,Hatefipour2022Aug} works. 
The characteristic property is the existence of chiral Andreev states generated when the chiral edge states of the quantum Hall system get in contact to the superconductor. In the filling factor $\nu=1$, there is a single chiral edge state. When it propagates in contact to a superconductor, it becomes a chiral Andreev state, which consists in an interference of particles and holes propagating chirally.  

In Ref.~\onlinecite{Panu2024Oct}, a two-dimensional electron gas in the quantum Hall state with $\nu=1$ is considered in a two-terminal configuration. 
A third terminal is defined by contacting to a superconductor with $s$-wave pairing $\Delta$ in a region of finite length. The magnetic field $B$ generating the quantum Hall effect is assumed to be perfectly screened into the superconductor. An electrical bias $eV$ is applied at one of the terminals of the quantum Hall system, while the other terminal as well as the superconductor are assumed to be grounded. 
The following conditions are assumed: (i) $eV, k_B T<\Delta$ so that the transport window is fully within the superconducting gap, (ii) 
$eV, k_B T < \Delta_B$ 
so that the Zeeman splitting is larger than the energy defining the transport window.
Separated electrical and thermal currents flow in response to the
electrical bias: a pure electrical current flows into the superconductor, while a pure heat current flows into the grounded terminal of the QHS. 
Intuitively, the fact that an electrical but no heat current flows into the superconductor can be understood by noticing that under the operational conditions no quasiparticle supragap transport is expected in the superconductor (ruling out the possibility of a heat current) while, because of Andreev reflection, a charge current is expected in the superconducting terminal~\cite{Mazza2015Jun}. 
Interestingly, the existence of a nonlinear heat current in the unbiased terminal of the QHS implies a nonlinear nonlocal Peltier effect in the device, despite the explicit fulfillment of the EIS in the linear regime. Notice that a two-terminal QHS under these operational conditions has a single pair of counter-propagating chiral edge states. 
This can be regarded as a perfect single-channel one-dimensional ballistic conductor with conductance $G=e^2/h$ and no thermoelectric response. Hence, the contact to the superconductor and the generation of a chiral Andreev state in this device enables the nonlocal and nonlinear thermoelectric response. Related all-normal effects require a barrier in the conductor to generate interferences~\cite{Sanchez2021Sep}. Additionally, thermal gradients in chiral edge states in the presence of superconductors may trigger intriguing $\Delta T$-noise properties~\cite{Pierattelli2025Jun}.

\subsection{Joule spectroscopy}
\label{sec:joule}
The excess current \cite{Tinkham2004Jun} in highly transmissive NS or SNS junctions provides a fingerprint for heating and cooling mechanisms in hybrid devices 
~\cite{Choi2010Jul, Tomi2021Oct,Ibabe2023May,Ibabe2024Jun}. The so-called ``Joule spectroscopy'' introduced in Ref. \onlinecite{Ibabe2023May} is based on this property. That work considered the case of SNS junctions defined on InAl nanowires fully covered by a thin epitaxial Al layer, which exhibit a hard induced superconducting gap~\cite{Chang2015Mar}. 

\begin{figure}
  \includegraphics[width=\linewidth]{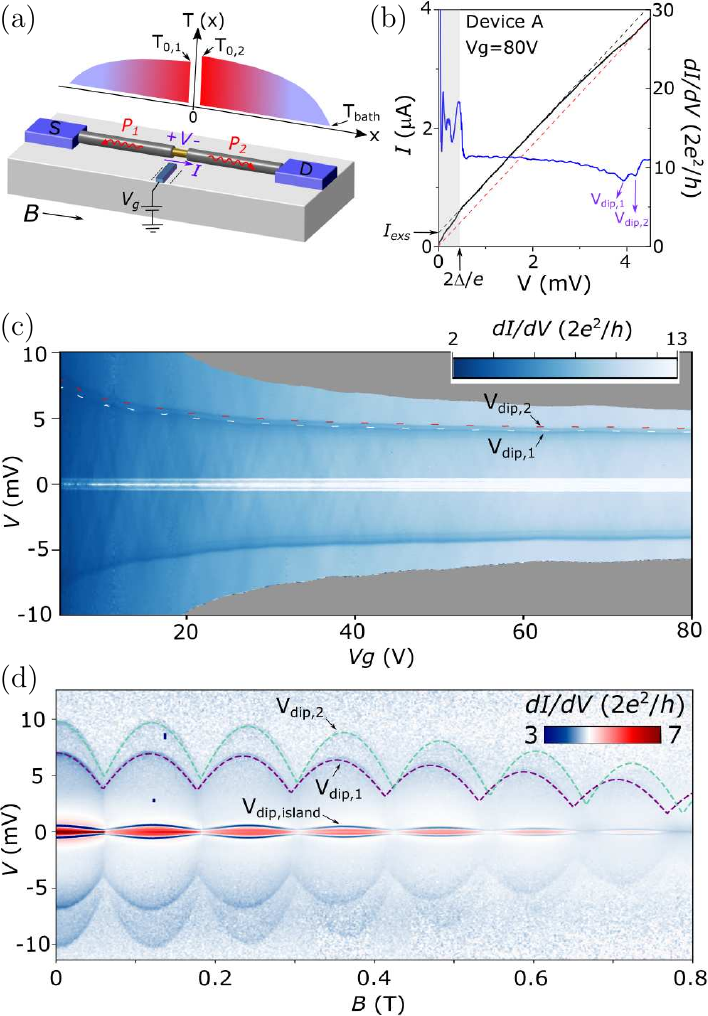}
  \caption{(a) Schematic of a highly transmitting SNS junction based on a hybrid superconductor-semiconductor nanowire. The temperatures of the S leads at the junction, $T_{0,i}$, can increase substantially due to the dissipated Joule power, $P_i$. (b) $V_{dip,i}$ are the voltages at which the S leads turn to the normal state due to Joule heating. They appear as dips in $dI/dV$, reflecting the suppression of excess current. (c) Gate dependence of $V_{dip,i}$. The dashed lines are fits to Eq.~\eqref{eq:joule-spectroscopy}, confirming the $\sqrt{R_J}$-dependence of the dips. (d) Joule spectroscopy of a device with a floating superconducting island and two grounded superconducting leads. In this case, three $dI/dV$ dips are observed, one for each superconductor-to-normal metal transition. The magnetic field-dependence of the dips of the floating and grounded superconductors are different, owing to their distinct dominant heat dissipation mechanisms. With courtesy of Ref.~\onlinecite{Ibabe2023May}.}
  \label{fig:joule-spectroscopy}
\end{figure}

The measured IV curves exhibited a series of differential conductance dips at voltages $V_{\rm dip, i} \gg 2\Delta$~\cite{Ibabe2023May}, which could be associated with the suppression of the excess current due to the transition to the normal state of certain portions of the device (the index $i=1,2$ corresponds to these different portions). The positions of the dips were found to be tunable as a function of several control parameters such as gate voltage, magnetic field and temperature, see Fig.~\ref{fig:joule-spectroscopy}. In particular, $V_{\rm dip, i}$ was found to scale as $\sqrt{R_J}$, where $R_J$ is the normal state junction resistance, tunable by means of a gate potential acting on the junction region.

The observed features and the working principle of the technique are described by a theoretical model that relies on the balance between the heat injected into the junction by the Joule effect and the different cooling mechanisms~\cite{Ibabe2023May} (i.e. quasiparticle diffusion and electron-phonon coupling). As these mechanisms are largely inefficient at low temperatures, the junction temperature increases with respect to the bath temperature of the cryostat $T_{\rm bath}$, which might eventually lead to a transition of the leads to the normal state.  
For a large bias voltage $V \gg \Delta$ the junction IV curve is well described as
\begin{equation}
    I = V/R_J + I_{\rm exc,1}(T_{0,1}) + I_{\rm exc,2}(T_{0,2}) \;,
    \label{IV-angel}
\end{equation}
where $I_{\rm exc,i}$ are the excess currents associated with lead $i$, and $T_{0,i}$ is the corresponding lead temperature, which can be different due to possible device asymmetries.  

In the single junction case~\cite{Ibabe2023May}, quasiparticle diffusion provides the main cooling mechanism. This allows to determine the power dissipated for $V_{\rm dip,i}$ as
\begin{equation}
P_{\rm dip,i} = \Lambda \frac{k^2_B T^2_{\rm c,i}}{e^2 R_{\rm lead,i}} \;,
\label{Pdip}
\end{equation}
where $T_{\rm c,i}$ is the critical temperature on lead $i$, $R_{\rm lead,i}$ is the lead normal resistance and $\Lambda \simeq 2.112$, which arises from the BCS thermal conductivity assuming $T_{\rm bath} \ll T_{\rm c,i}$. From Eqs. (\ref{IV-angel},\ref{Pdip}) and assuming a symmetric situation where $P_1 = P_2 \simeq V^2/2R_J$ one obtains
\begin{equation}
    V_{\rm dip,i} = \sqrt{\frac{2\Lambda R_J}{R_{\rm leads,i}}} 
    \frac{k_b T_{\rm c,i}}{e} \;,
\label{eq:joule-spectroscopy}
\end{equation}
relating the dips position to several device properties. 

Further devices that contain floating superconducting segments (islands), in addition to grounded superconducting leads (here the index $i = 1, 2, \text{island}$, where 1 and 2 refer to the leads) have been considered~\cite{Ibabe2024Jun}. In this case, it is shown that the cooling of islands by quasiparticle diffusion is strongly suppressed owing to the relatively high resistances of the SNS junctions. As a result, instead of Eq.~\eqref{Pdip}, the heat balance in islands is better described by an electron-phonon cooling term,
\begin{equation}
P_\text{dip,island} = \Sigma U \left(T_{c,\text{island}}^n - T_\text{bath}^n\right) \label{eq:PhCooling},
\end{equation}
where $\Sigma$ is a material-dependent parameter, $U$ is the volume of the island, and the exponent $n$ falls in the range 5-6, according to fits to the experimental data taken for devices based on InAs-Al nanowires.

\section{Two-dimensional Dirac materials}
\label{sec:2degs}

Dirac materials are a class of condensed-matter systems with linear low-energy band dispersion, nodal points, and high carrier mobility enabled by suppressed backscattering~\cite{Wehling2014Jan}. Despite differing compositions, their charge carriers behave like relativistic Dirac fermions, showing universal infrared responses such as in optical conductivity or specific heat. Protected by specific symmetries, Dirac nodes arise in materials ranging from 2D sheets (e.g., graphene, silicene, transition metal dichalcogenides) to topological insulator edge states, making them ideal platforms for nanoscale quantum devices.

Owing to the high carrier mobility and tunable electronic properties in Dirac materials, hybrid superconducting junctions based on them present promising avenues for thermoelectric applications. 
For example, in two-dimensional graphene-like systems, electrostatic gating near the Dirac point enables phenomena such as specular Andreev reflection~\cite{Beenakker2006Aug}, while gap engineering allows for optimization of thermoelectric performance. 
Moreover, low heat capacity and weak electron-phonon coupling in graphene enable sensitive bolometric detection using graphene-superconductor junctions~\cite{Vora2014Feb,Lee2020Oct,Vischi2020May}. 
Similarly, the spin-momentum-locked surface states of topological insulators hold significant potential for spintronic applications, offering efficient spin transport with suppressed backscattering~\cite{Xu2017Sep, Fu2020Apr}.

\begin{figure}[t]
  \includegraphics[width=\linewidth]{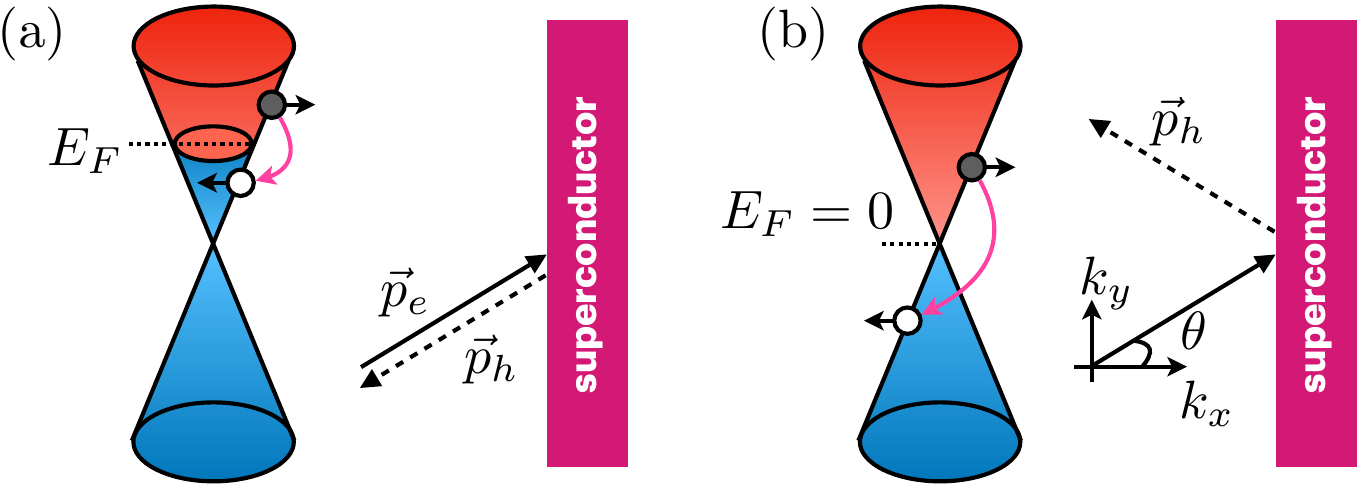}
  \caption{Andreev reflection in 2D Dirac materials. 
  (a) Andreev retro-reflections take place when $E_F>E$ and are dominant for $E_F\gg E,\Delta$. The incident electron (full circle) and reflected hole (empty circle) are both in the conduction band, and the hole retraces the trajectory of the electron. 
  (b) Specular Andreev reflection occurs for $E>E_F$ and dominates when $E_F\ll E,\Delta$. The reflected hole is in the valence band and its group velocity is parallel to its momentum. The reflection becomes specular since the momentum parallel to the interface ($k_y$) is conserved. }
  \label{fig:2d-dirac}
\end{figure}

Superconductivity in Dirac materials is usually induced by proximity to a conventional superconductor in a hybrid junction. 
We first consider the general Dirac Hamiltonian in two-dimensional space~\cite{Tamara2024May}
\begin{equation}\label{eq:general-dirac}
    \check{H}(x,y)= - \mi \hbar v_F \left( \check{\alpha}_x \partial_x + \check{\alpha}_y \partial_y \right) + \check{V}(x,y), 
\end{equation}
with $v_F$ the Fermi velocity, $\check{V}(x,y)$ the electrostatic potential, and the $2N\times 2N$ matrices $\check{\alpha}_{x,y}$ acting on the SU(2) degree of freedom defining the Dirac Hamiltonian and the $N$-dimensional space corresponding to the other degrees of freedom of the system. 
For example, for the surface state of a three-dimensional topological insulator (3dTI) the matrices $\check{\alpha}_{x,y}$ reduce to the Pauli matrices in spin space. By contrast, for graphene the Dirac Hamiltonian acts in sublattice space, denoted by A- and B-type atoms, but the general Hamiltonian \eqref{eq:general-dirac} also spans in the spin and valley degrees of freedom, becoming a $8\times8$ matrix ($N=4$). 

In the presence of superconducting correlations, the Dirac-Bogoliubov-de Gennes (DBdG) equations generalize \cref{eq:BdG} by including \cref{eq:general-dirac} as the single-particle Hamiltonian, and extending the pairing matrix to the appropriate space~\cite{Linder2008Feb,Herrera2010Jun,Burset2015Nov}. 
Graphene and other 2D graphene-like materials have two or a higher even number of Dirac nodes in their spectrum~\cite{CastroNeto2009Jan}. Usually called \textit{valleys}, these extra Dirac points add another degree of freedom to the DBdG equations. 
By contrast, the surface state of a 3DTI features a single Dirac point, and the DBdG Hamiltonian adopts its simplest form: a $4\times4$ matrix equation in electron-hole and spin space. 

To evaluate the transport properties of hybrid junctions between Dirac materials and superconductors, it is usually assumed that the contact width $W$ is much larger than the junction length $L$. 
Further assuming translational invariance along the contact direction, the momentum parallel to the interface between the normal and superconducting regions, $k_y$, is conserved, see \cref{fig:2d-dirac}. 
Under these approximations, the transverse momentum $k_y$ can be parametrized as the angle $
\sin\theta= k_y/\mathbf{k}_e
$, with $
\mathbf{k}_{e,h} = \hbar v_F |E\pm E_F|
$ the magnitude of the wavevector for electrons and holes, $v_F$ and $E_F$ respectively being the Fermi velocity and Fermi energy, and $E$ being the excitation energy. 
For an Andreev reflection process, the angle of the reflected hole is computed from the conservation of the transverse momentum as
\begin{equation}\label{eq:angle-h}
    \sin\theta_A = \frac{ E+E_F }{ |E-E_F| } \sin\theta . 
\end{equation}
Andreev reflections are thus only possible below a critical angle $
\sin\theta_c = \mathbf{k}_{h}/\mathbf{k}_{e}
$. 

The charge and heat currents, \cref{eq:Il,eq:Jl}, include the contributions from all transverse modes. For wide junctions, the sum over transverse modes results in the integral over the angle~\cite{Bercioux2018Dec}
\begin{equation}\label{eq:angle-avg}
    \frac{W}{2\pi}\int_{-\infty}^{\infty} \md k_y 
    = \frac{1}{2} N(E) \int_{-\pi/2}^{\pi/2} \md \theta \cos\theta ,
\end{equation}
with $N(E)= 2W|E+E_F|/(hv_F)$ the number of transverse modes. 
The factor $N(E)$ in \cref{eq:angle-avg} introduces an energy dependence unique to the two-dimensional problem involving Dirac materials. In the high-doping regime with $E_F\gg E$, the energy dependence of this prefactor becomes irrelevant and the angle-average resembles that of a conventional ballistic junction, see Sec.~\ref{sec:andreev}. By contrast, in the low-doping regime, the energy dependence of the number of transport channels creates interesting asymmetries for thermoelectric effects. 

In normal metals, the Fermi energy is much larger than the superconducting gap ($E_F \gg \Delta$), so Andreev reflection involves electrons and holes from the same band with nearly opposite momenta, see \cref{fig:2d-dirac}(a). This results in intra-band, retro-reflections. By contrast, Dirac materials allow for low-doping regimes where $E_F \lesssim \Delta$, enabling inter-band Andreev reflection: the reflected hole comes from the valence band and moves in the same direction as its momentum—specular reflection~\cite{Beenakker2006Aug}, as sketched in \cref{fig:2d-dirac}(b). 
While inter-/intra-band transitions can occur in 1D systems, see \cref{sec:wires}, retro-/specular Andreev reflection becomes meaningful in 2D, where they lead to unusual charge and heat transport, especially near the Dirac point.

\subsection{Graphene and graphene-like materials}

Graphene is a single layer of carbon atoms arranged in a honeycomb structure, constructed as the superposition of two triangular lattices denoted by A and B atoms. As a Dirac material, low-energy excitations in graphene follow a linearized Hamiltonian with two inequivalent valleys, which we assume to lie in the $x$-axis and denote them $\mathbf{K}_\pm = (\pm K,0)$~\cite{CastroNeto2009Jan}. 
Around each valley, single-particle excitations follow a Dirac Hamiltonian of the form~\cite{Linder2008Feb,Herrera2010Jun} 
\begin{align}
    \check{H}_\pm(\mathbf{k})={}& 
    \hat{h}_\pm(\mathbf{k}) 
    \hat{\sigma}_0 + \check{V}(x,y), \label{eq:graphene-dirac}
    \\
    \hat{h}_\pm(\mathbf{k}) ={}& \hbar v_F \left( \hat{s}_x k_x \pm \hat{s}_y k_y \right) , \label{eq:graphene-dirac-subl}
\end{align}
where the Pauli matrices $\hat{s}_{x,y,z}$ ($\hat{\sigma}_{x,y,z}$), with $\hat{s}_0$ ($\hat{\sigma}_0$) the identity matrix, act on sublattice (spin) space, and the velocity $v_F=\sqrt{3}t_g a/(2\hbar)\approx 10^6$~m~s$^{-1}$ is proportional to the nearest-neighbor hopping $t_g\approx 3$~eV and the lattice constant $a=0.246$~nm. 

In the vicinity of a superconductor, a proximized graphene region presents spin and valley degeneracy, allowing us to effectively consider only one valley for electrons and the opposite for holes. For proximized structures, it is also a common approximation to only consider pairing terms that are diagonal in sublattice space. Under these conditions, the DBdG Hamiltonian for graphene and graphene-like materials reduces to a $4\times4$ matrix in sublattice and Nambu (particle-hole) spaces, namely, 
\begin{equation}\label{eq:DBdG-graphene}
    \check{H}_\text{DBdG}(\mathbf{k}) = \begin{pmatrix}
       \hat{h}_\pm(\mathbf{k}) - E_F\hat{s}_0 & \Delta(\mathbf{k},T) \hat{s}_0 \\
       \Delta^*(\mathbf{k},T) \hat{s}_0 & E_F\hat{s}_0- \hat{h}_\pm(\mathbf{k})
    \end{pmatrix} ,
\end{equation}
with the sublattice Dirac Hamiltonians defined in \cref{eq:graphene-dirac-subl}. 
We are considering a general pairing which could be unconventional with an anisotropic momentum dependence, but in most cases we use isotropic $s$-wave pairings $\Delta(\mathbf{k},T)= \Delta(T)$. 
The self-consistent temperature dependence of the superconducting gap is commonly approximated by the interpolation formula
$\Delta(T) \approx \Delta_0 \tanh ( 1.74 \sqrt{T_c/T-1} )$,
with $T_c$ and $\Delta_0$ the proximity-induced critical temperature and the zero-temperature gap amplitude, respectively. 
The scattering probabilities required to compute the charge and heat currents, \cref{eq:Il,eq:Jl}, are obtained by matching the solutions of \cref{eq:DBdG-graphene} between normal ($\Delta_0=0$) and superconducting regions. These currents then need to be angle-averaged over all incident modes according to \cref{eq:angle-avg}. 

Early works already pointed out the important role that specular Andreev reflections could take in transport~\cite{Titov2007Jan}. For example, a lowly or undoped ($E_F\ll\Delta_0$) finite graphene region between two superconductors displays neutral boundary modes that propagate parallel to the normal-superconductor interfaces. These modes arise from sequential specular Andreev reflections at each interface, are sensitive to the phase difference between the superconductors, and carry energy but not charge. Consequently, the thermal conductance of such an Andreev channel becomes sensitive to the superconducting phase difference~\cite{Titov2007Jan,Liu2024Feb}. 

Another feature of Dirac materials is the conservation of chiral symmetry in transport, which leads to Klein tunneling effects in hybrid junctions~\cite{Allain2011Oct}. Transmission through a tunnel barrier between two Dirac materials is unusually high and oscillates with the product of the barrier height and width. 
This oscillatory behavior results in an unusual thermal conductance at graphene-based normal-superconductor junctions~\cite{Yokoyama2008Apr}. 
Additionally, a negative differential thermal conductance $\mathrm{d}J_N/\mathrm{d} (\delta T)<0$, with $\delta T$ being the thermal gradient, is also predicted for graphene NS junctions with singlet $s$-wave~\cite{Zhou2016May} and triplet $p$-wave superconductors~\cite{Liu2020Jul}. The unusual transport properties of Dirac materials coupled to superconductors results in a heat current that increases as the temperature gradient decreases, which can even be used to create or amplify a heat current~\cite{Li2006Apr}. 
Furthermore, electron cooling in a graphene sheet contacted by two insulator-superconductor junctions can outperform similar systems based on metallic ultra-thin films~\cite{Vischi2020May}. Graphene's low electron-phonon coupling preserves cooling mechanisms avoiding overheating, thus making graphene-based superconducting hybrid junctions interesting potential applications as bolometric radiation sensors~\cite{Vora2014Feb,Lee2020Oct,Vischi2020May}.  

Oscillations of the thermal conductance have also been predicted for graphene-based ferromagnetic-superconductor junctions~\cite{Salehi2010Jun,Salehi2010Oct,ZargarShoushtari2014Dec}. At low temperatures, the thermal conductance displays a minimum value around $h\sim E_F$, with $h$ being the exchange field from the ferromagnetic region. The temperature dependence of the thermal conductance is exponential for gaped superconductors~\cite{Yokoyama2008Apr,Salehi2010Jun,Salehi2010Oct}, that is, with $s$- or $d_{x^2-y^2}$-wave pairings. By contrast, nodal pairings like the $d_{xy}$-wave predicted at high-T$_c$ superconductors, display a linear behavior with the temperature, similar to the Wiedemann-Franz law for metals~\cite{Salehi2010Oct,ZargarShoushtari2014Dec}. Similar behavior is predicted for graphene-like materials like silicene~\cite{Paul2016Oct}, borophane~\cite{Zare2019Sep}, transition metal dichalcogenides~\cite{Majidi2014Oct, Majidi2022Jul}, or buckled 2D materials~\cite{Zhuo2023Sep}. 

The unusual specular Andreev reflection in low-doped graphene also leads to interesting thermoelectric effects. Seebeck thermopower is enhanced in the specular Andreev reflection regime of NS junctions~\cite{Wysokinski2013Apr}, while high values of charge and spin thermopower have been predicted for ferromagnetic-superconductor junctions with Rashba spin-orbit coupling~\cite{Beiranvand2017Feb}. 

Bilayer graphene is also a semimetal in its pristine form, although it no longer features bands with linear dispersion. However, the band gap of bilayer graphene is controllable by electrical gating, which allows for tunable thermoelectric effects in bilayer graphene-superconductor junctions~\cite{Bernazzani2023Apr}. 
The nonlinear bipolar thermoelectric effect induced by spontaneous electron-hole symmetry breaking is predicted to produce high Seebeck coefficients reaching up to $1\text{~mV}/\text{K}$ for bilayer temperatures of $T\sim10\text{~K}$, with surface power density of roughly $1 \text{~nW}/\mu$m$^2$ at maximal absolute thermodynamic efficiency of up to $40\%$~\cite{Bernazzani2023Apr}. 

\subsection{Edge states of 3D topological insulators}

The surface states of 3D topological insulators emerge from bulk spin-orbit coupling and band inversion and are another example of Dirac materials~\cite{Hasan2010Nov}. The spins of Dirac fermions on these surface states are rigidly connected to their momentum. Such intrinsic topological protection results in suppressed backscattering by impurities or defects and, consequently, in low-dissipation transport and spin currents that could be manipulated electrically. Topological insulators are thus ideal for energy-efficient electronics, and their thermoelectric properties have been explored in junctions without superconductors~\cite{Xu2017Sep, Fu2020Apr}. 

For hybrid junctions where superconductivity is proximity-induced on a surface state of a 3D topological insulator, the DBdG Hamiltonian is expressed as
\begin{equation}\label{eq:DBdG-3dti}
    \check{H}_\text{DBdG}(\mathbf{k}) = \begin{pmatrix}
       \hat{h}_0(\mathbf{k}) - E_F\hat{\sigma}_0 & i\hat{\sigma}_y\Delta(\mathbf{k},T)  \\
       -i\hat{\sigma}_y\Delta^*(\mathbf{k},T)  & E_F\hat{\sigma}_0- \hat{h}_0^*(\mathbf{-k})
    \end{pmatrix} ,
\end{equation}
which spans in spin ($\hat{\sigma}_{x,y,z}$ Pauli matrices) and Nambu spaces, and where $ 
\hat{h}_0(\mathbf{k}) = \hbar v_F \left( \hat{\sigma}_x k_x + \hat{\sigma}_y k_y \right) 
$ is the low-energy Dirac Hamiltonian. 
The pairing in \cref{eq:DBdG-3dti} is assumed to be singlet, either $s$- or $d$-wave. 

Cooling power in conventional NS junctions requires that the interface is in the tunnel limit~\cite{Bardas1995Nov}. 
Consequently, reduced backscattering from topological protection seems to be detrimental for heat transport in topological NS junctions. 
However, for highly-transparent NS junctions based on surface states of topological insulators, cooling power as a function of applied voltage can be important in the low-doping regime where specular Andreev reflections dominate~\cite{Bercioux2018Dec}. 
Anomalous thermal properties have been predicted when the superconductor has $d$-wave pairing symmetry~\cite{Ren2013Apr}. 
For example, the Kapitza resistance quantifies the thermal resistance at the interface between two dissimilar materials as heat flows across it. 
Topological NS junctions with $d$-wave pairing feature asymmetric Kapitza resistance, wherein the interface behaves as an efficient thermal conductor under a positive thermal bias, but as a poor conductor (or effective thermal insulator) under a negative bias~\cite{Ren2013Apr}.

The interplay between spin-singlet pairing from the superconductor and the spin-orbit coupling at the surface state leads to the emergence of spin-triplet pairings around the proximized region. 
The differential thermal conductance varies with the pairing state components, showing negative conductance only in spin-singlet dominated regimes~\cite{Ren2013Apr,Li2017Oct}. Spin-singlet states lead to phase- and length-dependent oscillations in conductance, which diminish as spin-triplet contributions increase~\cite{Li2017Oct}. This provides a potential method to identify pairing components in topological-insulator-based superconducting hybrids~\cite{Li2017Oct}. 

Another example of topological matter are Weyl semimetals, which are gapless in the bulk at discrete points and exhibit open surface states (Fermi arcs)~\cite{Armitage2018Jan}. 
The Weyl semimetal phase can be realized by breaking either time reversal or inversion symmetry. However, in the absence of a spin-active interface Andreev reflection is suppressed due to chirality blockade for time-reversal broken Weyl semimetal/Weyl superconductor junctions. 
As for other Dirac materials, the thermal conductance in the low-doping regime for inversion broken junctions features an exponential dependence on the temperature and thermopower with a high figure of merit around Weyl points~\cite{Saxena2023May}. Unusual thermal diode effects have also been predicted for inversion symmetry-broken Weyl superconductor/Weyl semimetal Josephson junctions~\cite{Chatterjee2024Jul}. 

\section{Conclusion}
\label{sec:concl}

We reviewed how heat and thermoelectrical transport are influenced by superconducting proximity effect on normal and topological materials, classifying the different phenomena based on the dimensionality of the electronic degrees of freedom. In nanoscale and multiterminal zero-dimensional geometries based on quantum dots, local and nonlocal features emerge as a consequence of nonlocal correlations induced by superconducting terminals. 
Special attention is given to how interactions and nonlocal superconducting correlations give rise to unconventional thermoelectric phenomena. 
Mesoscopic one-dimensional states emerge in quantum Hall regimes, 
where helical modes or spin-orbit coupling can induce distinctive heat transport and thermoelectric behavior. 
We also reviewed how non-equilibrium Joule heating in nanowire-based Josephson junctions would affect their Josephson properties. 
Finally, we examined how graphene-like 2D materials, and 3D topological and Weyl materials, may exhibit unconventional responses such as thermoelectricity, anomalous thermal transport, and thermal rectification when proximized by superconductors. 

The discussed cases and comprehensive survey of recent literature highlight the promise of hybrid superconducting systems for thermal control, novel functionalities, and applications in superconducting quantum technologies, taking the form of power generators or refrigerators that exploit or mitigate undesired hotspots, or of thermal diodes, transistors or switches able to operate with millikelvin temperature differences through nanoscale distances.

\medskip
\medskip
\textbf{Acknowledgements} \par 
We acknowledge funding from the Spanish Agencia Estatal de Investigaci\'on projects Nos.~PID2022-142911NB-I00, PID2020-117992GA-I00, PID2023-150224NB-I00 and CNS2022-135950, from the Spanish CM ``Talento Program'' project No.~2019-T1/IND-14088 and No.~2023-5A/IND-28927, and through the ``Mar\'{i}a de Maeztu'' Programme for Units of Excellence in R{\&}D CEX2023-001316-M.  
L.A. acknowledges support from CONICET, Argentina as well as FonCyT, Argentina, through Grant No.  PICT 2020-A-03661.
A.B. acknowledges the MUR-PRIN 2022—Grant No. 2022B9P8LN-(PE3)-Project NEThEQS “Non-equilibrium coherent thermal effects in quantum systems” in PNRR Mission 4-Component 2-Investment 1.1 “Fondo per il Programma Nazionale di Ricerca e Progetti di Rilevante Interesse Nazionale (PRIN)” funded by the European Union-Next Generation EU and CNR project QTHERMONANO. A.B. acknowledges discussions with A. Jordan, F. Taddei, M. Governale, F. Giazotto and the kind hospitality by the Institute of Quantum Studies at Chapman University where partially the work was done. 

\medskip



\bibliography{biblio.bib}

%
%
%
%

\end{document}